\documentclass[a4paper,11pt]{article}
\pdfoutput=1 

\usepackage{jcappub} 

\usepackage[T1]{fontenc} 

\usepackage{caption}
\usepackage{multirow}
\usepackage{tabstackengine}
\TABstackMath

\title{Implementation of frequency-correlated noise in CMB component separation: Method, Validation, and Early Applications}

\author[a,b,c,1]{L. Zapelli,\note{Corresponding author.}}
\author[b,c]{M. Bersanelli,}
\author[b,c]{L. P. L. Colombo,}
\author[b,c]{A. Mennella,}
\author[d]{L. G. Bachar,}
\author[e,f]{E. Battistelli,}
\author[g]{I. Bekkaoui,}
\author[h]{L. A. Bianchi,}
\author[e,f]{A. Coppolecchia,}
\author[i,j]{B. Costanza,}
\author[e,f]{P. De Bernardis,}
\author[e,f]{G. De Gasperis,}
\author[e,f,k,l]{S. Ferazzoli,}
\author[h]{M. Galloway,}
\author[d]{B. García,}
\author[m,n]{M. Gervasi,}
\author[o,p]{N. Mirón-Granese,}
\author[q]{J-Ch. Hamilton,}
\author[r]{T. Laclavère,}
\author[h]{J. G. S. Lunde,}
\author[e,f]{S. Masi,}
\author[g]{Y. Moulane,}
\author[s,t]{S. Paradiso,}
\author[d]{I. D. Queirolo,}
\author[r]{M. Regnier,}
\author[u]{G. E. Romero}
\author[i,j,v]{and C. G. Scóccola}

\affiliation[a]{Università di Trento -- CUP E66E23000110001,\\
Dipartimento di Fisica, Via Sommarive, 14, Trento, Italy}
\affiliation[b]{Dipartimento di Fisica, Università degli Studi di Milano,\\
Via Celoria, 16, Milano, Italy}
\affiliation[c]{INFN sezione di Milano,\\
20133 Milano, Italy}
\affiliation[d]{Instituto de Tecnologías en Detección y Astropartículas (CNEA, CONICET, UNSAM),\\Argentina
}
\affiliation[e]{Sapienza Università di Roma, Dipartimento di Fisica,\\
P.le A. Moro 5, Roma, Italy
}
\affiliation[f]{Istituto Nazionale di Fisica Nucleare -- Sezione di Roma1,\\
P.le A. Moro 5, Roma, Italy
}
\affiliation[g]{College of Physical Sciences and Engineering, Mohammed VI Polytechnic University,\\
43150, Ben Guerir, Morocco
}
\affiliation[h]{Institute of Theoretical Astrophysics, University of Oslo,\\
Blindern, Oslo, Norway
}
\affiliation[i]{Facultad de Ciencias Astronómicas y Geofísicas, Universidad Nacional de La Plata, Observatorio Astronómico,\\
Paseo del Bosque, B1900FWA La Plata, Argentina
}
\affiliation[j]{Consejo Nacional de Investigaciones Científicas y Técnicas (CONICET),\\
Rivadavia 1917, Buenos Aires, Argentina
}
\affiliation[k]{Università di Roma Tor Vergata, Dipartimento di Fisica,\\
Via della Ricerca Scientifica 1, Roma, Italy
}
\affiliation[l]{Istituto Nazionale di Astrofisica -- Osservatorio Astronomico di Cagliari,\\
Via della Scienza 5, Selargius (CA), Italy
}
\affiliation[m]{Università di Milano - Bicocca,\\
Milano, Italy
}
\affiliation[n]{INFN Milano - Bicocca,\\
Milano, Italy
}
\affiliation[o]{Universidad de Buenos Aires, Facultad de Ciencias Exactas y Naturales,\\
Departamento de Física, y CONICET
}
\affiliation[p]{Universidad de Buenos Aires, Instituto de Física de Buenos Aires (IFIBA),\\
Buenos Aires, Argentina
}
\affiliation[q]{Université Paris Cité, CNRS, Astroparticule et Cosmologie,\\
F-75013 Paris, France
}
\affiliation[r]{Univ Toulouse, CNES, CNRS, IRAP,\\
Toulouse, France
}
\affiliation[s]{Istituto Nazionale di Astrofisica -- Osservatorio di Astrofisica e Scienza dello Spazio di Bologna,\\
via Gobetti 101, I-40129 Bologna, Italy
}
\affiliation[t]{Istituto Nazionale di Fisica Nucleare, Sezione di Bologna,\\
viale Berti Pichat 6/2, I-40127 Bologna, Italy
}
\affiliation[u]{Instituto Argentino de Radioastronomía (IAR),\\
C. C. No. 5, 1894 Villa Elisa, Buenos Aires, Argentina
}
\affiliation[v]{Departamento de Física, FCFM, Universidad de Chile,\\
Blanco Encalada 2008, Santiago, Chile
}

\emailAdd{luca.zapelli@unitn.it}
\emailAdd{marco.bersanelli@fisica.unimi.it}
\emailAdd{loris.colombo@unimi.it}
\emailAdd{aniello.mennella@fisica.unimi.it}
\emailAdd{gomezlucajavier@gmail.com}
\emailAdd{elia.battistelli@roma1.infn.it}
\emailAdd{0000096270a9b9d5-dmarc-request@in2p3.fr}
\emailAdd{l.a.bianchi@astro.uio.no}
\emailAdd{alessandro.coppolecchia@roma1.infn.it}
\emailAdd{belen@fcaglp.unlp.edu.ar}
\emailAdd{paolo.debernardis@roma1.infn.it}
\emailAdd{giancarlo.degasperis@roma1.infn.it}
\emailAdd{sofia.ferazzoli@uniroma1.it}
\emailAdd{mathew.galloway@astro.uio.no}
\emailAdd{beatrizgarciautn@gmail.com}
\emailAdd{massimo.gervasi@unimib.it}
\emailAdd{nahuelmirongranese@gmail.com}
\emailAdd{hamilton@apc.in2p3.fr}
\emailAdd{laclavere@apc.in2p3.fr}
\emailAdd{j.g.s.lunde@astro.uio.no}
\emailAdd{silvia.masi@roma1.infn.it}
\emailAdd{Youssef.MOULANE@um6p.ma}
\emailAdd{simone.paradiso@inaf.it}
\emailAdd{ivan.queirolo@iteda.gob.ar}
\emailAdd{regnier@apc.in2p3.fr}
\emailAdd{gustavo.esteban.romero@gmail.com}
\emailAdd{claudiascoccola@gmail.com}

\abstract{
Parametric component separation methods used in cosmic microwave background (CMB) analyses commonly assume instrumental noise to be uncorrelated between frequency channels. This approximation may become insufficient for experiments whose design naturally introduces cross-channel correlations, such as interferometric or multi-band bolometric systems.
This work aims to provide a consistent framework to include frequency-correlated noise in \texttt{Commander2} and to quantify its impact on component separation in controlled and instrument-motivated simulations.
We generalized the amplitude sampling of the \texttt{Commander2} framework to consistently handle dense frequency-frequency noise covariance matrices, modifying the likelihood evaluation and parallel architecture. The implementation is validated on low-resolution simulations of a simplified CMB-S4-like configuration. We applied the generalized framework to instrument-motivated simulations of the QUBIC bolometric interferometer, using representative noise models derived from end-to-end simulations.
Validation tests show excellent convergence and agreement between sampled parameters and analytical maximum-likelihood estimates. Accounting for frequency correlations reduces the width of the maximum-likelihood residual distributions by up to 8\% on average, and up to 11\% for individual parameters, relative to the diagonal-noise approximation. In QUBIC-based simulations, accounting for an anti-correlated noise produces posterior widths reduced by about 60\%, while a positive-correlation test increases them by about 40\%.This shows that neglecting cross-channel correlations can lead to incorrect posterior estimates in either direction.
The generalized implementation presented here enables the  \texttt{Commander2} software to consistently propagate such correlations into parameter estimation, providing a framework for assessing their impact in future CMB analyses.
}


\arxivnumber{2609.04510}

\begin{document}
\maketitle
\flushbottom

\section{Introduction}\label{Sec1}
The Cosmic Microwave Background (CMB) is a nearly homogeneous radiation field that dates back to approximately 380,000 years after the Big Bang. Small anisotropies in its intensity and polarization encode key information about the early stages of the Universe and provide powerful constraints on cosmological models. 

Measuring the CMB requires separating the CMB from the other astrophysical emissions of different origin, called foregrounds. This process, called ``component separation'', exploits the distinct spectral behaviour of these components, which can be disentangled using data from multifrequency observations.

A wide variety of approaches have been developed for this task, such as blind separation techniques~\cite{a,b} and parametric methods~\cite{c,d} based on physically motivated models of the sky and noise. In particular, the parametric approach consists of modelling each emission component through a spectral and spatial template with free parameters that are fit directly from the data.

Parametric methods are widely used and highly effective, but current codes assume that instrumental noise is uncorrelated between
frequency channels. This simplification is primarily adopted for computational efficiency and is generally validated through dedicated simulation runs and cross spectra. 

In most CMB experiments, cross-detector correlations are typically found to be negligible with respect to statistical uncertainties~\cite{e}. When such correlations are found to be relevant, however, the response is often to exclude some subsets of data from the final analysis, or at least to restricting the kind of analyses that can be reliably performed.
An example is provided by the \textit{Planck} 2018 analysis~\cite{f,g}, where HFI half-ring maps were found to be affected by residual cross-correlations and therefore replaced with the odd-even maps, mitigating the issue at the cost of some loss of optimality.

For the level of accuracy required in the search for primordial B-mode polarization, such assumptions may no longer be sufficient~\cite{h}. This is particularly relevant for instruments employing interferometric or multi-band bolometric technologies, such as QUBIC~\cite{i}, LiteBIRD~\cite{j}, or future ground-based CMB-S4-like setups\footnote{The Cosmic Microwave Background Stage 4 (CMB-S4) project was officially discontinued during the preparation of this work. Its instrumental characteristics are nevertheless retained here as a representative high-sensitivity configuration for validation purposes.
}~\cite{k}, where non-negligible cross-channel correlations can naturally arise.
In such cases, neglecting cross-channel noise correlations may affect both the component reconstruction and the uncertainty estimation~\cite{l}.

In this context, we present an extension of the parametric component separation code \texttt{Commander2}~\cite{m,n} to incorporate a more general noise model that includes correlations between different frequency channels.
The implementation corresponds to replacing the usual block-diagonal noise covariance in frequency space in the map-domain likelihood with a covariance matrix that also contains cross-channel terms.
This approach is complementary to time-domain analyses, such as those performed within the \texttt{Commander3} framework~\cite{d}.

On the one hand, time-domain modelling can describe instrumental effects in greater detail, involving a substantially larger parameter space.
However, depending on the instrument and on the map-making procedure, the resulting maps may still contain non-negligible cross-channel noise correlations.
On the other hand, a map-domain approach allows one to investigate the impact of frequency correlations on component separation with a lower computational cost and without the need for a dedicated time-domain model for each experiment.

This simplification also defines the scope of this work. 
For instance, in the case of QUBIC, a full analysis would require a detailed treatment of the map-making process, including the modelling of the synthesized-beam peaks and their amplitudes. In the simulations considered here, we bypass this step and work directly with simulated maps and their associated covariance matrices. This allows us to isolate the effect of frequency-frequency noise correlations in a controlled setting.

The structure of the paper is as follows. In section~\ref{Sec2}, we describe the noise modelling framework and detail the implementation of frequency-correlated noise handling within the \texttt{Commander2} code, focusing on the necessary modifications to its data structures and computational architecture. In section~\ref{Sec3}, we outline the validation strategy and show results obtained from the upgraded \texttt{Commander2} on low-resolution simulations based on a toy model of a CMB-S4-like configuration. Here, CMB-S4-like refers to a representative high-sensitivity, multi-frequency ground-based CMB polarization setup, used only as a validation test case. We also show the effect of including frequency correlations using analytical computations. In section~\ref{Sec4}, we present an application of the extended code to simulated observations of the QUBIC experiment, based on an instrument-motivated configuration. 
The purpose of this analysis is not to provide a quantitative forecast, but rather to illustrate the practical relevance of frequency-correlated noise modelling in a physically relevant experimental setting based on QUBIC instrumental characteristics, and to assess the behaviour of the generalized implementation.
Finally, in section~\ref{Sec5}, we summarize our findings and discuss future directions, including the application of the code to more realistic simulations based on the official QUBIC pipeline~\cite{o,p}.

\section{Noise Model Extension}\label{Sec2}

\subsection{Component separation and the role of Noise Covariance}\label{sub2.1}
In parametric component separation, the observed sky signal in each frequency channel is modelled as a linear combination of astrophysical components plus instrumental noise. In its simplest form, the data model for a set of frequency maps can be written as
\begin{equation}\label{eq1}
    \mathbf{d}_\nu = \sum_c a_c(\nu)\,\mathbf{s}_c + \mathbf{n}_\nu \,,
\end{equation}
where $\mathbf{d}_\nu$ is the observed sky map at frequency $\nu$, $a_c(\nu)$ is the spectral response of component $c$ at frequency $\nu$, $\mathbf{s}_c$ is the map of component $c$ at a reference frequency, and $\mathbf{n}_\nu$ is the instrumental noise map at frequency $\nu$, assumed to be Gaussian. By stacking all frequencies and components, this model can be expressed in compact matrix notation as:
\begin{equation}\label{eq2}
    \mathbf{d} = \mathbf{A}\,\mathbf{s} + \mathbf{n} \,,
\end{equation}
where $\mathbf{A}$ is the ``mixing matrix'' encoding the spectral behaviour of each component across frequencies.

\texttt{Commander2} is a Bayesian parametric component separation code that estimates joint posterior distribution of the model parameters through a Gibbs sampling scheme, iteratively drawing samples from conditional distributions.
Within this framework, the component amplitudes $\mathbf{s}$ are sampled from the conditional posterior, proportional to the Gaussian likelihood
\begin{equation}\label{eq3}
    \mathcal{L}(\mathbf{d}|\mathbf{s}) \propto \exp\left[-\frac{1}{2} (\mathbf{d}-\mathbf{A}\mathbf{s})^T \,\mathbf{N}^{-1} \,(\mathbf{d}-\mathbf{A}\mathbf{s})\right] \,,
\end{equation}
where $\mathbf{N}$ is the noise covariance matrix (NCM), which stores all statistical properties of the instrumental noise. In pixel space, the most general $\mathbf{N}$ contains correlations:
\begin{enumerate}
    \item between different pixels (due to, e.g., scanning strategy);
    \item between Stokes parameters $I$, $Q$ and $U$;
    \item between different frequencies.
\end{enumerate}

Most existing implementations, including the original version of \texttt{Commander2}, assume that noise is uncorrelated across frequencies, which allows $\mathbf{N}$ to be block-diagonal. Although this is often justified by the given experiment noise characteristics, it does not generally hold in all cases. 

For the specific case of \texttt{Commander2}, this assumption was adopted for both architectural and scientific reasons. The code was primarily developed for the analysis of high-resolution full-sky satellite datasets, such as those provided by the \textit{Planck} and WMAP missions. In this context, even when considering only intra-channel correlations between map pixels and Stokes parameters, the full noise covariance matrix quickly becomes prohibitively large. 
When cross-channel correlations are included, the problem goes from scaling with the number of frequencies $N_\nu$ to scaling with $N_\nu \left(N_\nu + 1\right)/2$.
Although not strictly computationally unfeasible, this would significantly increase the complexity and cost of both memory storage and numerical operations, such as matrix inversion. For example, at HEALPix $N_\mathrm{side}=16$, a single frequency $\left(N_\mathrm{pix} \times N_\mathrm{Stokes}\right) \times \left(N_\mathrm{pix} \times N_\mathrm{Stokes}\right)$ covariance (with double precision floating point entries) occupies around 680 MB. For 10 frequency channels, the total size goes from 6.8 GB to 37.4 GB, following this general approach.

The level of frequency correlations in the noise of these instruments was assessed to be negligible within statistical uncertainties. Given this, the additional computational burden of a dense noise model was not considered justified. To give a numerical example, for WMAP the cross-correlation coefficient of different radiometers `1' and `2' was evaluated to be $C_{12}\approx0.007$ for the K band, and  $-0.001 < C_{12} < 0$ for the other bands~\cite{q}.
However, there are specific instrumental designs, like the one implemented in QUBIC, for which cross-channel correlations must be accurately modelled to avoid introducing systematic biases in the inferred component maps.

The goal of this work is to extend the \texttt{Commander2} noise model to support a fully general NCM including frequency-frequency correlations\footnote{Source code can be found at \href{https://github.com/LucaZapelli/freqcorr_commander2/}{https://github.com/LucaZapelli/freqcorr$\_$commander2/}.}. Formally, we allow $\mathbf{N}$ to be a dense matrix with non-zero off-diagonal blocks connecting different frequency channels:
\begin{equation}\label{eq4}
    \mathbf{N} = \begin{bmatrix}
                  \mathbf{N}_{\nu_1 \nu_1} & \mathbf{N}_{\nu_1 \nu_2} & \cdots\\
                  \mathbf{N}_{\nu_2 \nu_1} & \mathbf{N}_{\nu_2 \nu_2} & \cdots\\
                  \vdots & \vdots & \ddots
                 \end{bmatrix}\,.
\end{equation}
Each block $\mathbf{N}_{\nu_i \nu_j}$ represents the covariance between the noise in frequency channels $\nu_i$ and $\nu_j$. These blocks themselves are matrices encoding correlations over all pixels and Stokes components. When $\nu_i = \nu_j$, the block describes the auto-covariance of the noise in a single frequency channel, and when $\nu_i \neq \nu_j$, it describes the cross-channel correlations.
To use this general NCM in \texttt{Commander2}, significant changes to the code’s architecture and data flow were required.

\begin{figure}[t!]
    \centering
    \includegraphics[width=\linewidth]{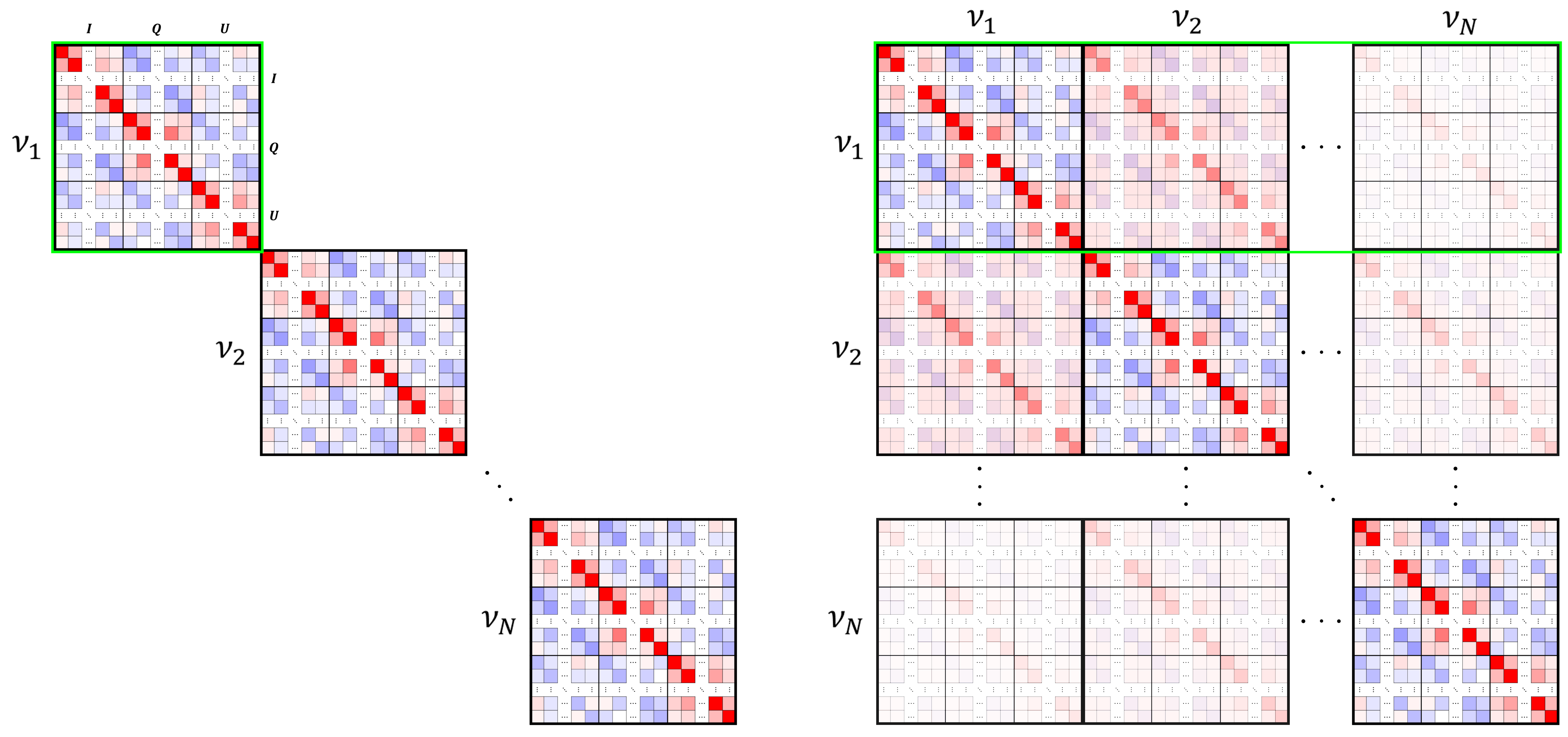}
    \caption{Comparison between the structure of the most general inverse noise covariance matrix used in the original \texttt{Commander2} code (left) and the one in the new implementation (right). For illustrative purposes, this example shows correlation matrices instead of covariances. The values are not intended to represent a specific experiment, as they are chosen only to highlight the internal structure and cross-channel formatting. In the original version of the code, each of the $N$ MPI processes $-$ assigned to one of the $N$ frequency channels $-$ loads in memory a $\left(N_\mathrm{pix} \times N_\mathrm{Stokes}\right) \times \left(N_\mathrm{pix} \times N_\mathrm{Stokes}\right)$ matrix, corresponding to its own channel. In the new implementation, each MPI process is still assigned to a primary frequency channel, but iteratively accesses and unloads the matrix blocks corresponding to an entire row (highlighted in green) of the full noise covariance, as needed.}
    \label{Fig1}
\end{figure}

\subsection{Parallel Architecture and Memory Handling}\label{sub2.2}
\texttt{Commander2} is designed for high-performance parallel execution on large datasets. In its original implementation, the code 
implicitly assumes a block-diagonal approximation of the noise covariance matrix in frequency space, which implies a factorization of the likelihood across frequency channels. Under this assumption, each MPI process handles the computations associated with a single frequency band. 

This architecture supports the efficient execution of frequency-local operations such as beam convolution, map projection, foreground modelling, and evaluation of the likelihood, under the assumption that noise is uncorrelated between frequency channels. This frequency-local design minimizes the memory footprint per process and intra-process communication: each MPI task needs to load only the instrumental and signal data relevant to its own assigned band, including the bandpass response, beam profile, gain factors, signal maps, noise RMS maps, and (or) pixel-IQU-dense inverse NCMs.

The introduction of a fully dense noise covariance model, which includes frequency-frequency correlations, breaks the assumption of likelihood separability and therefore requires changes to the architecture of the code. In the generalized implementation, each MPI process still has a primary frequency assignment but must now participate in global operations that couple data across all channels (figure~\ref{Fig1}). In particular, operations involving the inverse noise covariance matrix $-$ such as the application of $\mathbf{N}^{-1} \mathbf{d}$, where $\mathbf{d}$ is the full multi-frequency data vector $-$ now require contributions from all frequency bands.
To accommodate this, each process loads into memory the full set of signal and instrument model information across all channels. This includes component templates, frequency bandpasses, and beam profiles. The implications in terms of memory requirements are discussed in appendix~\ref{Appendix A}.
By contrast, the inverse noise covariance matrix itself is not fully loaded into memory. Instead, its individual block components $\left(\mathbf{N}^{-1}\right)_{\nu_i\nu_j}$ are loaded from the row corresponding to the primary frequency band $\nu_i$, as needed during the computation.
These blocks, each of size $\left(N_\mathrm{pix} \times N_\mathrm{Stokes}\right)\times\left(N_\mathrm{pix} \times N_\mathrm{Stokes}\right)$, are structurally identical to the inverse auto-covariance blocks used in the original code and are loaded on demand at runtime to avoid excessive memory usage.
In the presence of cross-channel correlations, noise-weighted residuals must include contributions from all bands. 

For a process associated with frequency $\nu_i$, the contribution to the likelihood becomes:
\begin{equation}\label{eq5}
    \chi^2_{\nu_i} = (\mathbf{d}_{\nu_i}-\mathbf{A}_{\nu_i} \mathbf{s}) ^T\sum_{\nu_j} \left(\mathbf{N}^{-1}\right)_{\nu_i\nu_j} (\mathbf{d}_{\nu_j}-\mathbf{A}_{\nu_j} \mathbf{s})\,.
\end{equation}
In the absence of cross-channel correlations, where $\mathbf{N}^{-1}_{\nu_i\nu_j} = 0$ for $\nu_i \neq \nu_j$, this reduces to the standard per-channel chi-square expression, with $\chi^2_{\nu_i}$ representing an independent contribution to the overall likelihood. However, once correlations are included, this interpretation no longer holds: the expression above may yield negative values, since the cross-channel terms can contribute with either sign, and therefore does not correspond to a proper chi-square test statistic. Only the total quantity,
\begin{equation}\label{eq6}
    \chi^2 = \sum_{\nu_i} \chi^2_{\nu_i} \,,
\end{equation}
represents a valid measure of model fit. This generalization ensures that the influence of cross-channel noise correlations is consistently propagated into the likelihood and parameter estimation.

The new implementation supporting cross-frequency noise is currently limited to amplitude sampling only. Extending the same treatment to non-linear parameters is, in principle, straightforward. However, in practice, this leads to a substantial increase in computational cost.
In the original \texttt{Commander2} code, spectral parameters are sampled using an inverse transform sampling algorithm. This approach is particularly efficient, as it produces independent samples without the need for tuning. Moreover, the posterior distributions of spectral parameters are often significantly non-Gaussian, making inverse sampling especially well suited for this problem.
When cross-frequency correlations are included, each likelihood evaluation involves a significant I/O overhead due to the distributed structure of the covariance. As inverse sampling requires multiple likelihood evaluations to construct the underlying probability density and cumulative distribution functions, the computational cost becomes prohibitive.

Alternative approaches such as Metropolis–Hastings could reduce the number of likelihood evaluations per step. However, in this context, they typically suffer from lower acceptance rates and increased autocorrelation, leading to a comparable or higher computational cost per effective independent sample.

As for the memory requirements, the computational performance of the code in terms of wall-clock times is discussed in appendix~\ref{Appendix A}.

\section{Code Validation}\label{Sec3}

To validate the \texttt{Commander2} noise model extension described in section~\ref{Sec2}, we developed a validation framework structured in two phases, each corresponding to an instrumental and observational simulation setup.
In each phase, we assessed how the inclusion of cross-channel noise correlations in the likelihood affects the fidelity of component separation by comparing analytical estimates and sampled posterior distributions of the model parameters under matched sky conditions.

\subsection{Simulated Sky and Instrument Configuration}\label{sub3.1}
For validation purposes we simulated a CMB-S4-like toy model experiment observing a region covering 1.5\% of the sky, centred at the Galactic longitude $\ell = 316.45^\circ$ and latitude $b = -58.76^\circ$, corresponding to the QUBIC observation field centre to ensure consistency with the sky patch adopted in the QUBIC-based simulations presented in section~\ref{Sec4}. The mock instrument features nine frequency channels centred at 20, 30, 40, 85, 95, 145, 155, 220, and 270 GHz, each modelled with a top-hat bandpass response (see table~\ref{tab1}). For simplicity, no beam smoothing is applied to the sky maps. Since the validation runs are performed at very low resolution ($N_\mathrm{side}=4$), the instrumental beam is subdominant with respect to the pixel window function. Moreover, the analysis is based on relative comparisons between NCM configurations applied to the same data model, so common beam effects are not expected to significantly affect the results. 

The noise covariance matrix used to generate the data is dense in frequency and Stokes space, while spatial (pixel-pixel) correlations are set to zero. 
As a result, the component separation analysis was performed independently at each pixel. 
By excluding spatial correlations, we eliminated an additional source of mixing between reconstructed amplitudes that could complicate the interpretation of residuals and posterior widths. Although coupling between Stokes parameters remains present, the absence of spatial correlations allows a clearer isolation of the influence of frequency correlations, which is the specific aspect under investigation in this validation study.

The simulated noise model includes a representative level of 30\% positive correlation between frequency channels and Stokes parameters, chosen to be sufficiently strong to induce measurable effects while preserving a stable covariance structure.

The sky signal is composed of CMB and thermal dust emission. 
Synchrotron radiation is not included in this validation setup, although it is a relevant foreground component in real CMB observations, especially below 100 GHz. The goal of this analysis is not to reproduce a fully realistic sky model but to isolate the impact of noise covariance mismodelling on parameter reconstruction. Limiting the sky model to two components reduces additional foreground degeneracies and allows a cleaner comparison between NCM configurations.
We generated the CMB realizations using the CAMB library~\cite{s}, with the set cosmological parameters shown in table~\ref{tab2}, and the dust emission as that of a modified blackbody with the following spectrum,
\begin{table}[t]
\centering
  \renewcommand{\arraystretch}{1}
  \begin{tabular}{|c|c|c|c|c|}
   \hline
    \begin{tabular}[c]{@{}c@{}}$\nu$ \\ (GHz) \end{tabular}  & \begin{tabular}[c]{@{}c@{}}$\Delta\nu$ \\ (GHz) \end{tabular}  & \begin{tabular}[c]{@{}c@{}}$\sigma_I$ \\ ($\mu$K$\,$arcmin) \end{tabular} & \begin{tabular}[c]{@{}c@{}}$\sigma_{Q,U}$ \\ ($\mu$K$\,$arcmin) \end{tabular} \\\cline{1-4}
    20 & 5 & 16.5 & 14.93 \\
    30 & 9 & 9.36 & 8.52 \\
    40 & 12 & 11.85 & 10.79 \\
    85 & 20.4 & 2.02 & 1.85 \\
    95 & 22.8 & 1.78 & 1.63 \\
    145 & 31.9 & 3.89 & 2.52 \\
    155 & 34.1 & 4.16 & 2.70 \\
    220 & 48.4 & 10.15 & 6.58 \\
    270 & 59.4 & 17.4 & 11.29 \\ \cline{1-4}
  \end{tabular}
  \caption{Instrumental specifications of the CMB-S4-like experiment used in the validation runs. Listed from left to right: central frequency of each channel, bandpass width, white noise level in intensity, and white noise level in polarization~\cite{r}.}\label{tab1}
\end{table}
\begin{table}[t]
\centering
  \renewcommand{\arraystretch}{1.2}
  \begin{tabular}{|c|c|c|c|c|c|c|c|}
  \hline
    $\mathrm{H}_0$ & $\mathrm{\Omega}_b h^2$ & $\mathrm{\Omega}_c h^2$ & $\mathrm{\Omega}_K$ & $m_\nu$ & $\tau$ & $A_s$ & $n_s$\\ \cline{1-8}
    67.5 & 0.022 & 0.122 & 0 & 0.06 & 0.06 & $2 \times10^{9}$ & 0.965 \\ \cline{1-8}
  \end{tabular}
  \caption{Set of cosmological parameters used to define the CAMB angular power spectrum.}\label{tab2}
\end{table}
\begin{equation}\label{eq7}
    \mathrm{S}_\mathrm{d}(p,\nu) = \mathrm{s}_\mathrm{d}(p) \frac{e^\frac{kT_\mathrm{d}}{h\nu_{0,\mathrm{d}}}-1}{e^\frac{kT_\mathrm{d}}{h\nu} -1} \left( \frac{\nu}{\nu_{0,\mathrm{d}}} \right)^{\beta_\mathrm{d}+1}\,,
\end{equation}
after the d0 model of the PySM library~\cite{t,u,v}. 

This model assumes spatially constant temperature and spectral index, with values $T_\mathrm{d} = 20\,K$ and $\beta_\mathrm{d}=1.54$ respectively. The reference frequency for the dust component is set to $\nu_{0,\mathrm{d}} = 270\,$GHz.

Noise realizations are drawn from a zero-mean Gaussian process with dense covariance across frequencies and Stokes parameters, as described above. 
Per-channel intensity and polarization noise levels are listed in table~\ref{tab1}.


\begin{figure}[t!]
    \centering
    \includegraphics[width=\linewidth]{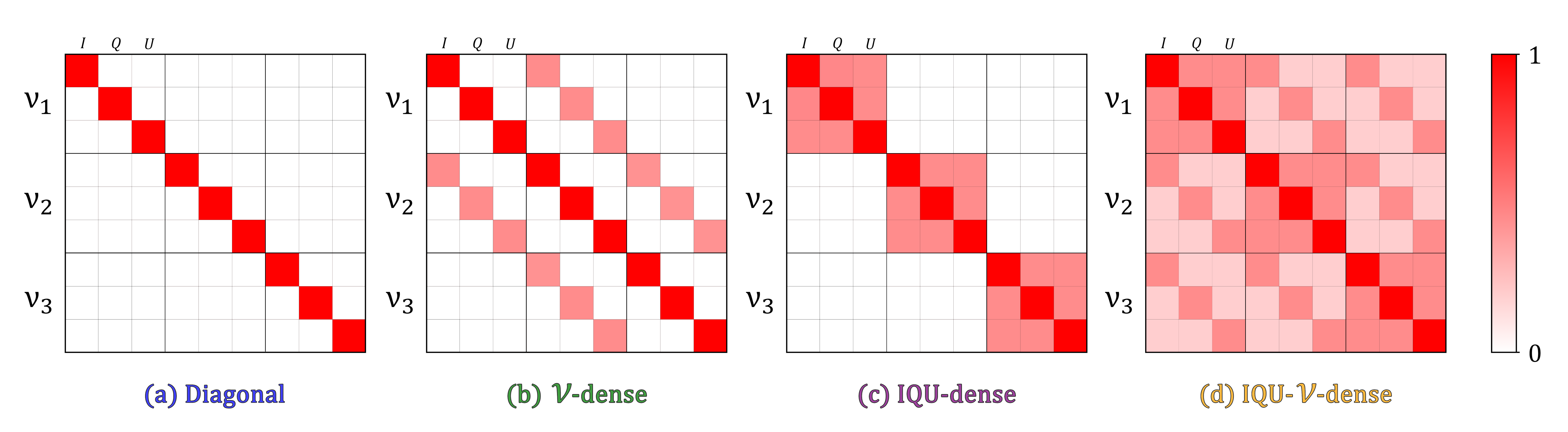}
    \caption{Example of the four NCM approximations adopted in the validation analysis, illustrated for a simplified instrumental setup with three frequency channels. The plots show correlation matrices (rather than covariances) for a single pixel, in order to visually emphasize the structural differences. From left to right, the panels represent: (a) the diagonal configuration $-$ uncorrelated frequency channels and Stokes components; (b) the $\nu$-dense configuration $-$ frequency-correlated, but no Stokes correlations; (c) the IQU configuration $-$ correlated Stokes components, but no frequency correlation; (d) the fully dense configuration $-$ correlations across both frequency channels and Stokes components.}
    \label{Fig2}
\end{figure}

\subsection{Validation Framework}\label{sub3.2}

The core of the validation lies in assessing the impact of different approximations to the noise covariance matrix. We defined four configurations of increasing complexity (figure~\ref{Fig2}):
\begin{enumerate}
    \item \textbf{Diagonal}: only the diagonal elements of the noise covariance matrix are retained. This implies complete independence between frequency channels and Stokes parameters. This configuration is color-coded in \textbf{blue} for the rest of the paper;

    \item $\boldsymbol{\nu}$\textbf{-dense}: frequency-frequency correlations are included, while Stokes correlations are neglected. This acts as a $\nu$-generalized extension of the diagonal case. This configuration is color-coded in \textbf{green};

    \item \textbf{IQU-dense}: Stokes parameter correlations are included, while frequency correlations are neglected. This is the most general form supported by the original \texttt{Commander2} implementation. This configuration is color-coded in \textbf{purple};

    \item \textbf{IQU-}$\boldsymbol{\nu}$\textbf{-dense} (or \textbf{fully dense}): both frequency and Stokes parameter correlations are preserved. This corresponds to the actual noise covariance used to generate the noise in the simulations and represents the most complete model tested. Note that this acts as a $\nu$-generalized extension of the IQU-dense configuration. This configuration is color-coded in \textbf{orange}.
\end{enumerate}

Throughout this work, we refer to the diagonal and the IQU-dense configurations as ``legacy'', since they can also be analysed with the original \texttt{Commander2} implementation. 
All simulated noise realizations in this work are generated using the IQU-$\nu$-dense covariance matrix. The other configurations are used only at the inference stage, as approximations to the true covariance.
The purpose of comparing a legacy configuration to its $\nu$-generalized extension is to isolate the impact of cross-channel noise correlations on parameter estimation. 
Since all other modelling assumptions are held fixed across comparisons, any observed differences can be directly attributed to the role of frequency correlations.

\subsubsection{Phase I - Correlations between neighbouring channels}\label{sub3.2.1}

In this stage, we carried out simulations at $N_\mathrm{side} = 4$, performing component separation on the two independent pixels that make up the sky patch. This choice maximized the signal-to-noise ratio, preserving a minimal map-based data structure. It also ensures that differences between legacy and $\nu$-generalized configurations can be more directly attributed to the treatment of frequency correlations rather than to limitations in component separation arising from poor foreground separability.
Since pixel-pixel noise correlations were not included and component separation was performed on independent pixels, increasing the resolution would mainly correspond to lowering the signal-to-noise ratio, rather than testing a qualitatively different aspect of the covariance model . A higher-resolution analytical check at $N_\mathrm{side}=256$ is discussed in section~\ref{sub3.3} to verify the behaviour of the same effect with a lower signal-to-noise.

\begin{figure}[!t]
    \centering
    \includegraphics[width=0.827\linewidth]{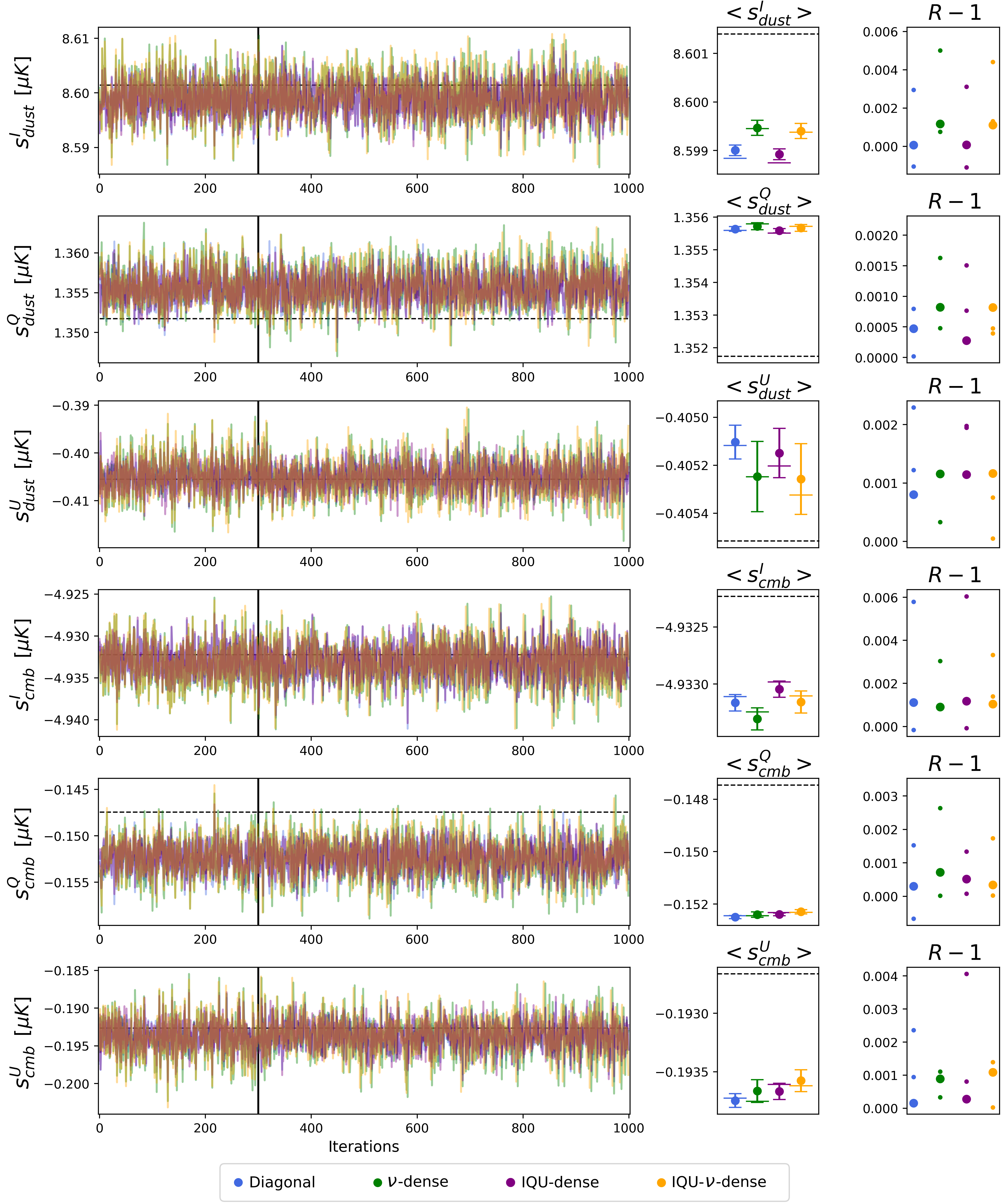}
    \caption{Phase I: 
    Each horizontal panel, consisting of three subplots, corresponds to one of the six sampled sky signal parameters $-$ three Stokes components ($I$, $Q$, $U$) for dust and CMB. The left subplot in each panel shows the evolution of the Markov chains obtained from a component separation run on a single noise realization, one for each NCM configuration (color-coded as described in section~\ref{sub3.2}), for one pixel. The black vertical line marks the end of the burn-in period (300 iterations); only the subsequent 700 samples are used in the statistical analysis. The central subplot in each panel summarizes the mean of the post-burn-in posteriors, with the related standard error. The horizontal segments represent the analytical maximum-likelihood estimates for the corresponding parameter, noise realization and NCM configuration. The horizontal black dashed lines 
    indicate the true values used for simulating the input maps.
    The right subplots display the Gelman–Rubin convergence statistic $R - 1$, computed from both the full (big marker) and half (small marker) chains, for each NCM configuration.}
    \label{Fig3}
\end{figure}

We generated the noise injected into the simulated maps from a IQU-$\nu$-dense covariance containing frequency correlations for neighbouring channels only, setting to zero the correlations between channels $\nu_i$,$\nu_j$ for which $|i-j|>1$. We generated a total of 100 independent noise realizations running five parallel sampling chains to assess convergence and internal agreement for all NCM configurations, as shown in figure~\ref{Fig3}. Every chain is composed of 1000 iterations, 30\% of which is discarded as burn-in.

The analysis proceeded in two main steps:
\begin{enumerate}
    \item \textbf{Chain Convergence}: we evaluated the convergence of sampling chains using the Gelman-Rubin diagnostic~\cite{w,x}, applying this test to each parameter to verify whether the chains reached a stable state to ensure statistically reliable results;
    \item \textbf{Agreement with Analytical Maximum-Likelihood}: here we compared the mean of each sampling chain to the analytical maximum-likelihood estimate (MLE) for the corresponding pixel and noise realization. Discrepancies are used to identify potential biases or numerical artifacts. We also verified that the new code recovers the results of the original implementation on the legacy configurations, i.e. in the absence of frequency correlations.
    \end{enumerate}

\paragraph{Chain Convergence}\label{par3.2.1.1}
To assess the convergence and reliability of the Markov Chain Monte Carlo (MCMC) sampling, we performed an independent Gelman–Rubin convergence diagnostic on each set of sampling chains (right panels of figure~\ref{Fig3}).
The Gelman–Rubin statistic $R$ evaluates whether multiple chains initialized from dispersed starting points have converged to the same target distribution. The test is based on comparing the variance within each chain with the variance between the chains.
For all sampled parameters and noise realizations, we obtained $R - 1 < 0.005$, which we consider sufficient to assess convergence on both pixels.\footnote{As a reference, \textsc{BeyondPlanck} set a threshold of 0.01 for cosmological parameter constraints~\cite{y}.}

To further reinforce the robustness of the convergence assessment, we also applied the same diagnostic separately to the first and second halves of each chain. This serves as a consistency check: the split chains independently satisfy $R - 1 < 0.012$ for all sampled parameters, thus strengthening the confidence that the chains have not only converged globally but also remain stable throughout the sampling process. This additional test helped identify potential non-stationarity or long autocorrelation times that might not have been apparent when evaluating the full chains alone.

\paragraph{Agreement with Analytical Maximum-Likelihood}\label{par3.2.1.2}
In addition to verifying convergence of the sampling chains, we assessed whether the MCMC estimates recover the expected MLE of the simulations. Since the likelihood is Gaussian, the MLE of each parameter can be computed analytically by solving a linear system of this form:
\begin{equation}\label{eq14}
    \hat{\mathrm{s}}_\mathrm{MLE} = \mathop{\arg\max}_\mathrm{s} \mathcal{L}(\mathbf{d} |\mkern 2mu \mathrm{s},\mathbf{N}^{-1}) \,.
\end{equation}

The MCMC estimates are expected to fluctuate around the analytical MLEs. The goal of this test is therefore to verify that the numerical uncertainty associated with the sampling is negligible compared to the width of the posterior distributions.

For each noise realization, we computed the difference between the mean values of the sampling chains and their corresponding analytical MLEs, 
\begin{equation}\label{eq15}
    \Delta\mathbf{s} = \bar{\mathbf{s}} - \hat{\mathbf{s}}_\mathrm{MLE} \,.
\end{equation}
In the rest of the paper, we refer to this quantity as the \textbf{MLE residual}.
Then, for each chain we estimated the integrated autocorrelation time
\begin{equation}\label{eq16}
    \tau = 1 + \sum_{k=1}^\infty \gamma_k \,,
\end{equation}
where $\gamma_k$ denotes the lagged autocorrelation function of the chain~\cite{z,aa,ab}. The average integrated autocorrelation length is small, less than 2 for all parameters, since the samples are drawn directly from the conditional posteriors.

We built our test statistic as 
\begin{equation}\label{eq17}
    \mathbf{z} = \mathbf{\Sigma}^{-1/2} \cdot \sqrt{\frac{N_\mathrm{iter}}{\mathbf{\tau}}} \cdot \Delta\mathbf{s}\,,
\end{equation}
where $\mathbf{\Sigma}$ is the unbiased sample covariance of the 6 parameters from the chains of a single run, to account for correlations between sky components.

\begin{figure}[t]
    \centering
    \includegraphics[width=\linewidth]{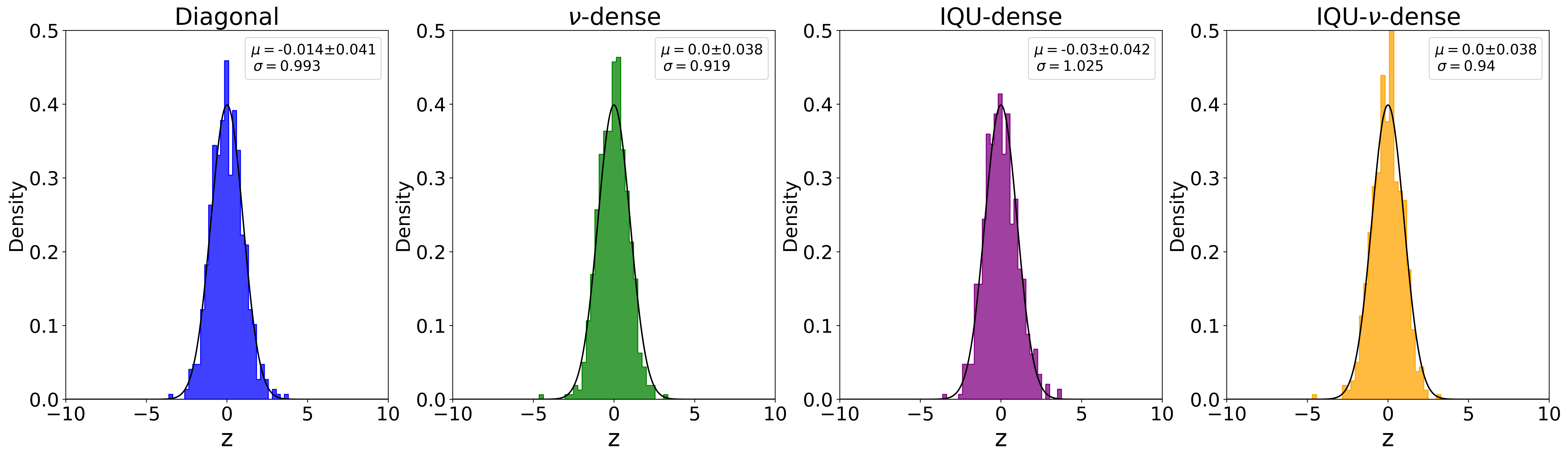}
    \caption{Phase I: Distributions of the test statistic defined in eq.~\ref{eq17} for each NCM configuration on a single pixel. Each histogram gathers statistics from all parameters and noise realizations. The black line shows a $\mathcal{N}(0,1)$ standard normal distribution. All objects are density-normalized to 1.}
    \label{Fig4}
\end{figure}

For properly sampled Gaussian posteriors, this random variable is expected to follow a normal distribution $\mathcal{N}(0, 1)$\footnote{The $z$ test statistic is actually distributed according to a Student's $t$ distribution with $N_\mathrm{eff}-1=N_\mathrm{iter}/\tau-1$ degrees of freedom. Since the number of effective iterations in a chain is much greater than 1, the distribution can be approximated with a standard normal distribution.}.
We tested this hypothesis by computing the p-value independently for all noise realizations and sampled parameters. The percentages of rejected chains at 95\% confidence level for different NCM configurations are the following (for both pixels): 5.8\% and 7.2\% for the diagonal configuration, 5.7\% and 3.7\% for the $\nu$-dense, 6.8\% and 7.1\% for the IQU-dense, and 6.5\% and 4.8\% for the IQU-$\nu$-dense one.
The distributions of the test statistic for the four NCM configurations are shown in figure~\ref{Fig4}. 
From these results, we concluded that the sampling procedure recovers the analytical maximum-likelihood solutions within the statistical limits of the adopted setup.

\subsubsection{Phase II - Correlations between all channels}\label{sub3.2.2}

The second phase of simulations was performed generating maps containing noise that is 30\% positively correlated across all nine frequency channels. The rest of the adopted setup is the same as Phase I. As shown in the appendix~\ref{Appendix A}, the computing time for the $\nu$-generalized configurations increased with respect to Phase I. This is due to the more complex structure of the cross-channel covariance blocks, which led the CG solver to take more iterations to reach the chosen residual tolerance.

Computing the Gelman-Rubin diagnostic, we obtained $R-1<0.004$ on all parameters and noise realizations for the full chains, and $R-1<0.012$ for the half chains.

The percentages of rejected chains at 95\% confidence level for both pixels are, respectively, 6.8\% and 7.0\% for the diagonal configuration, 2.2\% and 3.2\% for the $\nu$-dense, 7.3\% and 7.6\% for the IQU-dense, and 3.2\% and 2.8\% for the IQU-$\nu$-dense one.

\begin{table}[t]
\centering
\renewcommand{\arraystretch}{1.3}
\begin{tabular}{cc|c|c|c|c|c|c|c|}
\cline{3-9}
  &  & $\mathrm{s}_\mathrm{dust}^\mathrm{I}$ & $\mathrm{s}_\mathrm{dust}^\mathrm{Q}$ & $\mathrm{s}_\mathrm{dust}^\mathrm{U}$ & $\mathrm{s}_\mathrm{cmb}^\mathrm{I}$ & $\mathrm{s}_\mathrm{cmb}^\mathrm{Q}$ & $\mathrm{s}_\mathrm{cmb}^\mathrm{U}$ & \begin{tabular}[c]{@{}c@{}}Average ratio \\ over parameters \end{tabular} \\ \cline{1-9}

  \multicolumn{1}{|c|}{} & $\sigma_\mathrm{\nu}/\sigma_\mathrm{diag}$ & 0.98 & 0.99 & 0.98 & 0.97 & 0.98 & 0.98 & 0.98 \\
  \multicolumn{1}{|c|}{Phase I} & $\sigma_\mathrm{IQU}/\sigma_\mathrm{diag}$ & 1.00 & 1.00 & 1.00 & 1.00 & 1.00 & 1.00 & 1.00 \\
  \multicolumn{1}{|c|}{} & $\sigma_\mathrm{full}/\sigma_\mathrm{diag}$ & 0.98 & 0.99 & 0.98 & 0.97 & 0.98 & 0.98 & 0.98 \\ \cline{1-9}

  \multicolumn{1}{|c|}{} & $\sigma_\mathrm{\nu}/\sigma_\mathrm{diag}$ & 0.90 & 0.93 & 0.91 & 0.92 & 0.92 & 0.92 & 0.92 \\
  \multicolumn{1}{|c|}{Phase II} & $\sigma_\mathrm{IQU}/\sigma_\mathrm{diag}$ & 1.00 & 1.00 & 1.00 & 1.00 & 1.00 & 1.00 & 1.00 \\
  \multicolumn{1}{|c|}{} & $\sigma_\mathrm{full}/\sigma_\mathrm{diag}$ & 0.90 & 0.93 & 0.91 & 0.92 & 0.92 & 0.92 & 0.92 \\ \cline{1-9} 

  \multicolumn{1}{|c|}{} & $\sigma_\mathrm{\nu}/\sigma_\mathrm{diag}$ & 0.89 & 0.94 & 0.95 & 0.96 & 0.94 & 0.91 & 0.93 \\
  \multicolumn{1}{|c|}{Phase III} & $\sigma_\mathrm{IQU}/\sigma_\mathrm{diag}$ & 1.00 & 1.00 & 1.00 & 1.00 & 1.00 & 1.00 & 1.00 \\
  \multicolumn{1}{|c|}{} & $\sigma_\mathrm{full}/\sigma_\mathrm{diag}$ & 0.90 & 0.94 & 0.94 & 0.96 & 0.94 & 0.91 & 0.93 \\ \hline   
\end{tabular}
\caption{Ratios between the widths of residual distributions $\Delta\mathbf{s}_\mathrm{MLE}$ computed from analytical MLEs on all parameters. In particular, the ratios show the width of distributions from dense-correlation NCM configurations with respect to the diagonal case. Last row displays the averages over sky component amplitudes.} 
\label{tab4}
\end{table}

\subsection{Effects of Frequency Correlations}\label{sub3.3}

We investigated the impact of taking into account noise frequency correlations on the estimation of model parameters. The effect was studied on two quantities. First, we compared the MLEs with the corresponding input values. This calculation was performed analytically, without running the component separation. The second quantity is the joint posterior distribution of the parameters, as sampled through the component separation.
We describe the two analyses in detail below.

We computed the distribution of analytical MLEs under different noise covariance configurations. We examined how the choice of the NCM model affected both the bias and spread of the recovered parameters relative to the true input sky signal across many realizations. Specifically, we generated 500 simulated observations of the sky using the same configuration as in Phase I, and 500 realizations as in Phase II. We also generated 500 realizations at a higher resolution, namely $N_\mathrm{side} = 256$, studying the analytical results for a single pixel at the centre of the sky patch. This third set of maps, here called Phase III for continuity, contained noise that was correlated between all channels, as in Phase II. Since in all simulations the pixel-pixel correlations were set to zero, this high-resolution analysis serves only to check if the results hold for a lower signal-to-noise ratio.

\begin{figure}[t]
    \centering
\includegraphics[width=\linewidth]{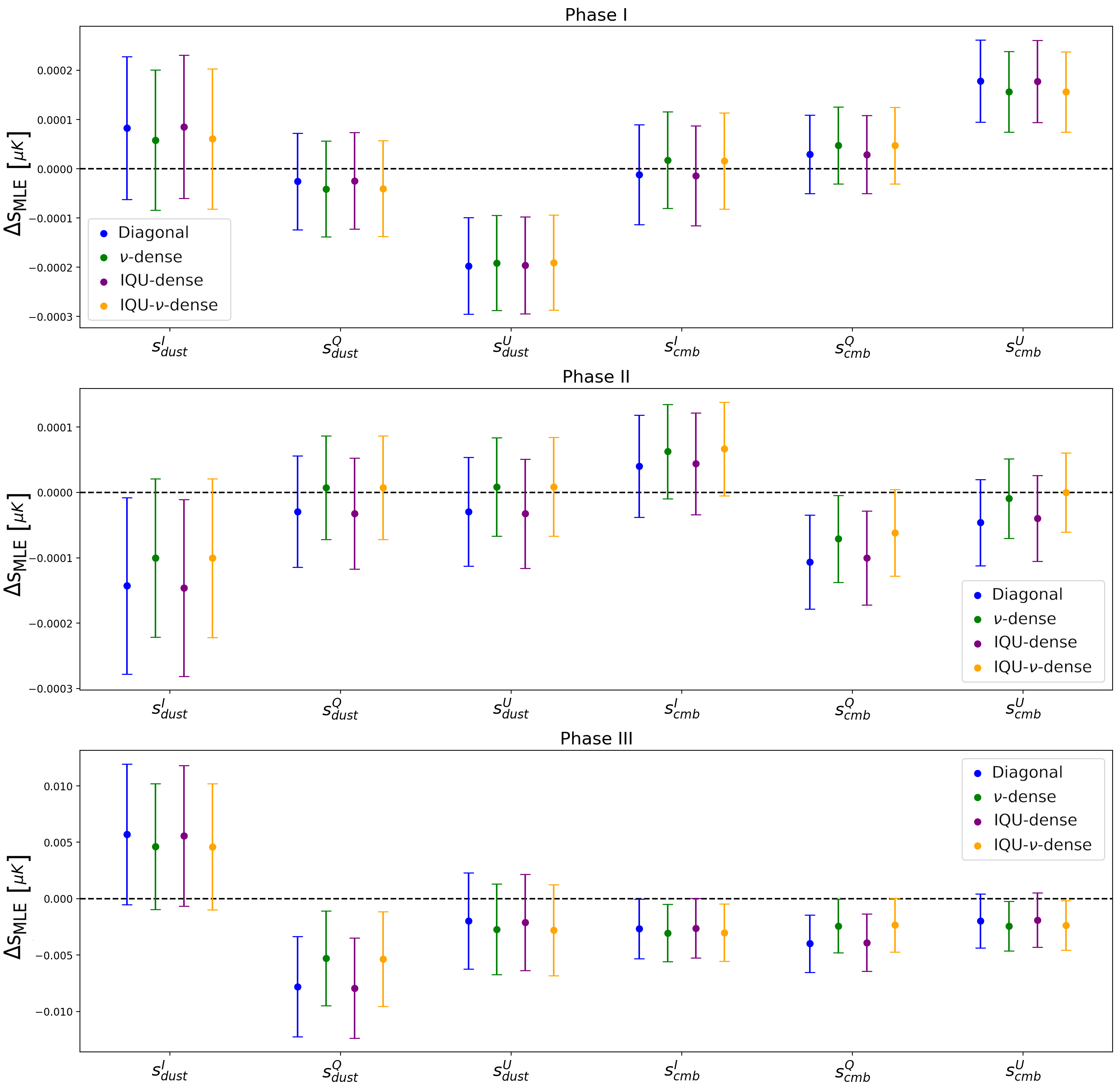}
    \caption{Mean values, with the corresponding standard errors on the mean, of the residual distributions computed from analytical MLEs. Results are divided by parameters and Phases, or instrumental configuration. Phase I and II simulations were performed at $N_\mathrm{side}=4$, injecting noise in the maps with a frequency correlation level of 30\% between neighbouring channels only (Phase I) and all channels (Phase II). Phase III maps were at $N_\mathrm{side}=256$, with noise correlated between all frequency channels.}
    \label{Fig5}
\end{figure}

For each noise realization, we defined the quantity
\begin{equation}\label{eq18}
    \Delta\mathbf{s}_\mathrm{MLE} = \hat{\mathbf{s}}_\mathrm{MLE} - \mathbf{s}_\mathrm{true} \,
\end{equation}
as the difference between the array of amplitude MLEs and the true input values.
%
%
%
%
%
%
%
We then built the distribution of these quantities by gathering the results from all noise realizations.
By comparing the distributions between NCM configurations, we can evaluate whether accounting for cross-channel correlations improves parameter recovery, particularly in terms of reducing statistical bias or uncertainty.
Table~\ref{tab4} displays the ratios of the distribution widths between configurations with dense covariance matrices and the diagonal case, showing that MLE residual distributions from $\nu$-generalized configurations are systematically more compact around the true input value compared to the ones from their legacy counterparts.
Figure~\ref{Fig5} shows the distribution means and their associated standard error for each parameter. Considering the cases for which the means of the legacy distributions are not compatible with the input value within $1\,\sigma$, the $\nu$-generalized configurations mitigate the discrepancy most of the time, and in several cases eliminate it.
The figure also shows that configurations sharing the same treatment of frequency correlations (i.e. diagonal and IQU-dense, or $\nu$-dense and fully dense) tend to give similar results, especially in Phase II and III. This result is expected. The nominal correlation level is the same for Stokes and frequency correlations (30\%), but the number of coupled maps is different. IQU correlations couple the three Stokes maps within each frequency channel, while frequency correlations couple a number of frequency maps depending on the considered Phase, from three to nine. Their effective impact in Phases II and III is therefore larger.

\begin{figure}[]
    \centering
    \includegraphics[width=\linewidth]{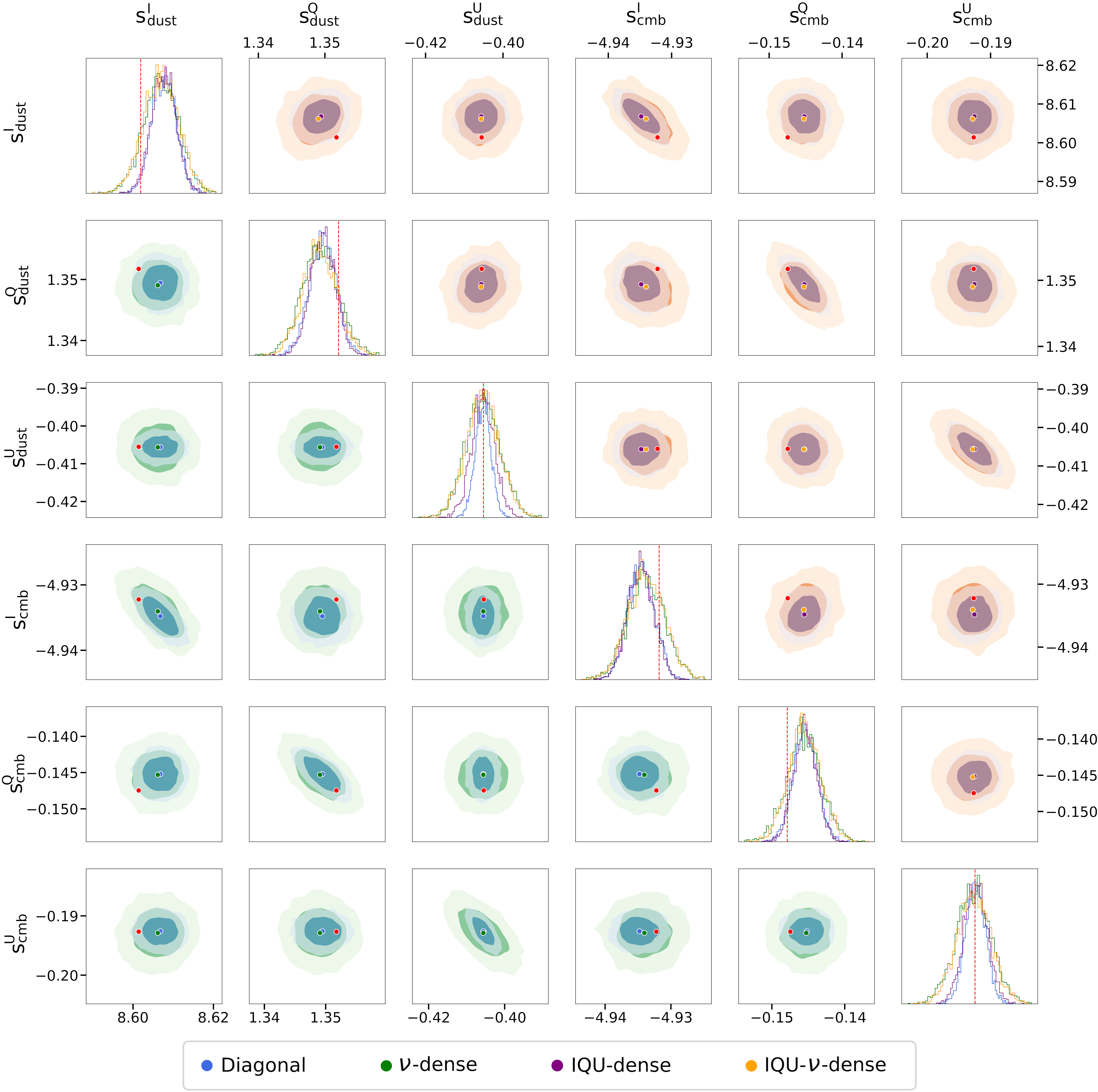}
    \caption{Phase I: Comparison of joint and marginal posterior distributions across NCM configurations from a single noise realization and pixel. The bottom-left corner plot displays the joint posterior distributions of the sampled parameters for the diagonal configuration (blue) and its $\nu$-generalized counterpart (green). The shaded contours combine samples from five parallel chains of the same noise realization, and enclose the central 68.2\% and 95.4\% intervals, respectively, going from the dark region to the light one. Coloured dots represent the analytical MLEs, while the red dots indicate the true input parameter values used in the simulations. The top-right corner plot shows the corresponding results for the IQU-dense configuration (purple) and its $\nu$-generalized case (orange). The histograms along the diagonal show the marginal posterior distributions for all NCM configurations, with vertical red lines marking the true input values}
    \label{Fig6}
\end{figure}
\begin{figure}[]
    \centering
    \includegraphics[width=\linewidth]{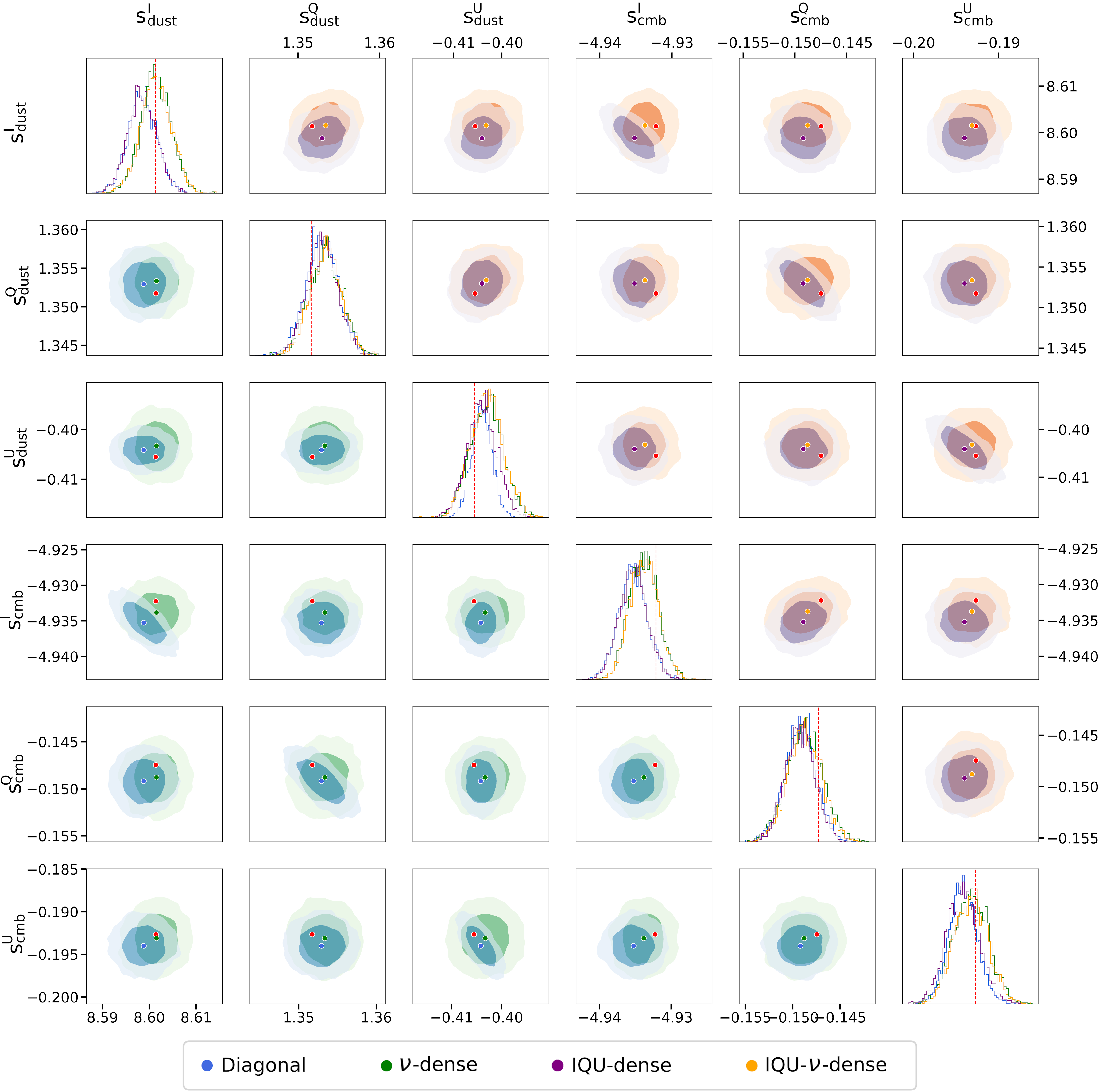}
    \caption{Phase II: Comparison of joint and marginal distributions from a single noise realization and pixel. This plot mirrors the one shown in figure~\ref{Fig6}. The joint posterior distributions are shown for the diagonal and IQU-dense NCM configurations (bottom-left and top-right corner, respectively in blue and purple), along with their $\nu$-generalized counterparts (green and orange). All contours and histograms are constructed from samples across five parallel sampling chains. Coloured dots represent the corresponding analytical MLEs, while red dots and vertical lines indicate the true input parameter values used in the simulations.}
    \label{Fig7}
\end{figure}

The second quantity that we studied is the joint posterior distribution of the sampled parameters.
In figures~\ref{Fig6} and~\ref{Fig7} we show the \texttt{Commander2} joint posterior distributions of the amplitudes for both Phase I and II, displaying them in two corner plots, each one comparing the results from a legacy NCM configuration and its $\nu$-generalized counterpart.
The results from the latter distributions visually show a higher frequency of true values falling within the central 68\% interval compared to the legacy configurations. In addition to that, degeneracies between parameters are generally reduced.
This effect is consistent for all noise realizations.

\section{Application to QUBIC-Motivated Simulations}\label{Sec4}

QUBIC provides a particularly relevant test case for the generalized implementation because its spectral imaging approach reconstructs multiple frequency sub-bands from measurements within the same physical wide band, naturally producing non-negligible correlations, as discussed below.
The simulations presented in this section are based on the detailed end-to-end characterization of QUBIC noise covariance published in~\cite{ac}. However, they are intended as an instrument-motivated application and a stress test of the generalized noise model extension, rather than as a forecast of QUBIC performance.
In particular, the covariance models used here retain frequency-frequency and Stokes-Stokes correlations, but neglect pixel-pixel noise correlations by construction. This choice allows us to isolate the impact of cross-frequency correlations on the component separation, without mixing this effect with additional correlations induced by the scanning strategy, synthesized beam, or map-making procedure. We also note that atmospheric noise is not included in the simulations presented in this section. For ground-based experiments, residual atmospheric fluctuations may produce significant correlations both between frequency channels and between pixels. Such pixel-pixel correlations are expected to be relevant in a full QUBIC analysis, but their characterization would require dedicated end-to-end simulations. 

More generally, this analysis illustrates how frequency-frequency noise correlations may affect component separation in experiments where multiple channels are reconstructed from common data, or show significant correlated noise contributions.

\subsection{QUBIC and Bolometric Interferometry}\label{sub4.1}
The Q \& U Bolometric Interferometer for Cosmology (QUBIC) is a ground-based experiment designed to measure CMB polarization. This is currently the only experiment based on a novel technique called bolometric interferometry~\cite{i}.

A bolometric interferometer collects the incoming radiation by means of an array of antennas, and projects it on a focal plane populated by bolometers, which operate measuring the total power of the signal over a single wide frequency band.
The combined optical response of the antennas results in an interference pattern, characterized by multiple peaks, and whose detailed shape depends on the specific pattern of the array. In particular, the angular separation of the peaks of this so-called ``synthesized beam'' is linearly dependent on the wavelength, as shown in~\cite{ac}. Consequently, the spatial distribution of the measured signal also contains spectral information, thus enabling the implementation of the so-called ``spectral imaging'' technique. This consists of decomposing a single wide frequency band into multiple narrower sub-bands in post-processing. The number of sub-bands can be chosen during data analysis, with an upper constraint given by the geometry of the array of antennas.
In particular, a regular $N_\mathrm{h} \times N_\mathrm{h}$ square grid of antennas on a side leads to a frequency resolution of $f_\mathrm{res}=\frac{\Delta\nu}{\nu}\simeq\frac{1}{N_\mathrm{h}-1}$~\citep{ac}.

While this capability is a major advantage in component separation for controlling foreground contamination~\cite{ad}, it also introduces complexity in the structure of the noise. Since the sub-bands are not independent physical channels but rather come from a partitioning of the same wide band, their noise contributions are inherently correlated. 

In QUBIC's specific case, end-to-end simulations show that noise in nearby sub-bands is significantly anti-correlated. This makes it an especially relevant test case for the methodology developed in this work.

\subsection{Simulated Instruments}\label{sub4.2}

As QUBIC is the first experiment of its kind, a Technological Demonstrator (TD) was built first to validate the technique. This is a preliminary version of the Full Instrument (FI) which will be used in the actual observation campaign that will lead to the main scientific results. In particular, the TD is characterized by a single wide frequency channel centred at 150 GHz, while the FI is capable of measuring with two independent wide bands centred at 150 and 220 GHz. 
Other main differences between the two instrumental configurations concern the focal planes (248 detectors for the TD, $992\times2$ for the FI) and the antenna arrays (64 horns for the TD, 400 horns for the FI), which provide a frequency resolution of, respectively, $f_\mathrm{res}^\mathrm{TD}\approx0.143$ and $f_\mathrm{res}^\mathrm{FI}\approx0.053$. These are the minimal fractional bandwidths that can be resolved through spectral imaging.
The TD is currently operative and collecting data, with the aim of verifying that the technique is working correctly. 

We simulated three instrumental configurations based on TD and FI specifics given in relevant QUBIC papers~\cite{i,ac,ae}. The details are given in table~\ref{tab5}.

\begin{table}[b!]
\centering 
  \renewcommand{\arraystretch}{1.2}
 \begin{tabular}{|c|c|c|c|}
\hline
Instrument name  & TD-3  & FI-1  & FI-3 \\ \hline
QUBIC configuration  & TD & FI & FI \\ \hline
\begin{tabular}[c]{@{}c@{}}Full band central\\ frequency $\nu_\mathrm{c}$ (GHz)\end{tabular}
 & \begin{tabular}[c]{@{}c@{}}150\\ -\end{tabular}                     & \begin{tabular}[c]{@{}c@{}}150\\ 220\end{tabular}            & \begin{tabular}[c]{@{}c@{}}150\\ 220\end{tabular}                                                    \\ \hline
 \begin{tabular}[c]{@{}c@{}}Full band bandpass\\ width $\Delta \nu$ (GHz)\end{tabular}
& 38   & \begin{tabular}[c]{@{}c@{}}38\\ 55\end{tabular}              & \begin{tabular}[c]{@{}c@{}}38\\ 55\end{tabular}                                                      \\ \hline
 \begin{tabular}[c]{@{}c@{}}Full band angular\\ resolution $\theta$ (arcmin) \end{tabular}
& 40.8   & \begin{tabular}[c]{@{}c@{}}23.4\\ 16.2\end{tabular}              & \begin{tabular}[c]{@{}c@{}}23.4\\16.2\end{tabular}                                                      \\ \hline
Number of sub-bands                                          & 3                                                                   & \begin{tabular}[c]{@{}c@{}}1\\ 1\end{tabular}                & \begin{tabular}[c]{@{}c@{}}3\\ 3\end{tabular}                                                        \\ \hline
\begin{tabular}[c]{@{}c@{}}Sub-band central\\ frequency $\nu_\mathrm{sub}$ (GHz)\end{tabular}                                & \begin{tabular}[c]{@{}c@{}}136.8\\ 148.9\\ 162.1\\ \\ \\ -\\ \\\end{tabular} & \begin{tabular}[c]{@{}c@{}}\\ 150\\ \\ \\ \\ 220\\ \\\end{tabular}   & \begin{tabular}[c]{@{}c@{}}136.8\\ 148.9\\ 162.1\\ \\ 200.9\\ 218.5\\ 237.6\end{tabular} \\ \hline
\begin{tabular}[c]{@{}c@{}}Sub-band bandwidth\\ $\Delta \nu_\mathrm{sub}$ (GHz)\end{tabular}           & \begin{tabular}[c]{@{}c@{}}11.6\\ 12.6\\ 13.8 \\ \\ \\ -\\ \\\end{tabular}    & \begin{tabular}[c]{@{}c@{}}\\ 38\\ \\ \\ \\ 55\\ \\\end{tabular}     & \begin{tabular}[c]{@{}c@{}}11.6\\ 12.6\\ 13.8\\ \\ 16.8\\ 18.3\\ 19.9\end{tabular}       \\ \hline
Fractional bandwidth $f_\mathrm{sub}$           & \begin{tabular}[c]{@{}c@{}}0.085\\ 0.085\\ 0.085 \\ \\ \\ -\\ \\\end{tabular}    & \begin{tabular}[c]{@{}c@{}}\\ 0.253\\ \\ \\ \\ 0.204\\ \\\end{tabular}     & \begin{tabular}[c]{@{}c@{}}0.085\\ 0.085\\ 0.085\\ \\ 0.084\\ 0.084\\ 0.084\end{tabular}       \\ \hline
Observing time $t_\mathrm{obs}$ (yr)   & 1  & 3        & 3 \\ \hline
$\sigma_{I}$ ($\mu \mathrm{K}\,\mathrm{arcmin}$)   & \begin{tabular}[c]{@{}c@{}}6793 \\ -  \end{tabular}  & \begin{tabular}[c]{@{}c@{}}9.7\\ 6.7\end{tabular}        & \begin{tabular}[c]{@{}c@{}}16.8\\ 11.6\end{tabular}                                              \\ \hline
$\sigma_{Q,U}$ ($\mu \mathrm{K}\,\mathrm{arcmin}$)   & \begin{tabular}[c]{@{}c@{}}9607  \\ -  \end{tabular} & \begin{tabular}[c]{@{}c@{}}13.7\\ 9.5\end{tabular}       & \begin{tabular}[c]{@{}c@{}}23.8\\ 16.4\end{tabular}                                              \\ \hline
\end{tabular}
\caption{Specifics of the simulated QUBIC instrumental configurations.}
\label{tab5}
\end{table}

\subsubsection{Band Partitioning and Frequency Resolution}\label{sub4.2.2}
To simulate an instrument-compatible partitioning of the wide bands for the considered configurations, we defined the $N_\mathrm{sub}$ sub-band central frequencies and bandwidths by solving the following system of equations:
\begin{equation}\label{eq21}
\setstackgap{L}{22pt}
\left\{\tabbedCenterstack[l]{
\sum\limits_{i=1}^{N_\mathrm{sub}}\Delta\nu_\mathrm{sub}^i=\Delta\nu=\nu_1-\nu_0
\\
\Delta\nu_\mathrm{sub}^i\cdot\left(1-f_\mathrm{sub}^i/2\right) = f_\mathrm{sub}^i\cdot(\nu_0+\sum\limits_{j=1}^{i}\Delta\nu_\mathrm{sub}^j)
\\
f_\mathrm{sub}^i=\nu_\mathrm{sub}^i/\Delta\nu_\mathrm{sub}^i\geq f_\mathrm{res}/2
}\right.\,.
\end{equation}
The set of equations in the last line of the system reflects the intrinsic frequency resolution of the instrument. In a bolometric interferometer, frequency information is contained in the spatial structure of the interference pattern, so a frequency band can be partitioned only if its fractional width is large enough to resolve different spectral contributions on the focal plane. The geometry of the array of antennas therefore sets the minimum scale $f_\mathrm{res}$: once the resulting sub-bands approach this scale, they cannot be further partitioned.

\begin{figure}[t]
    \centering
    \includegraphics[width=\linewidth]{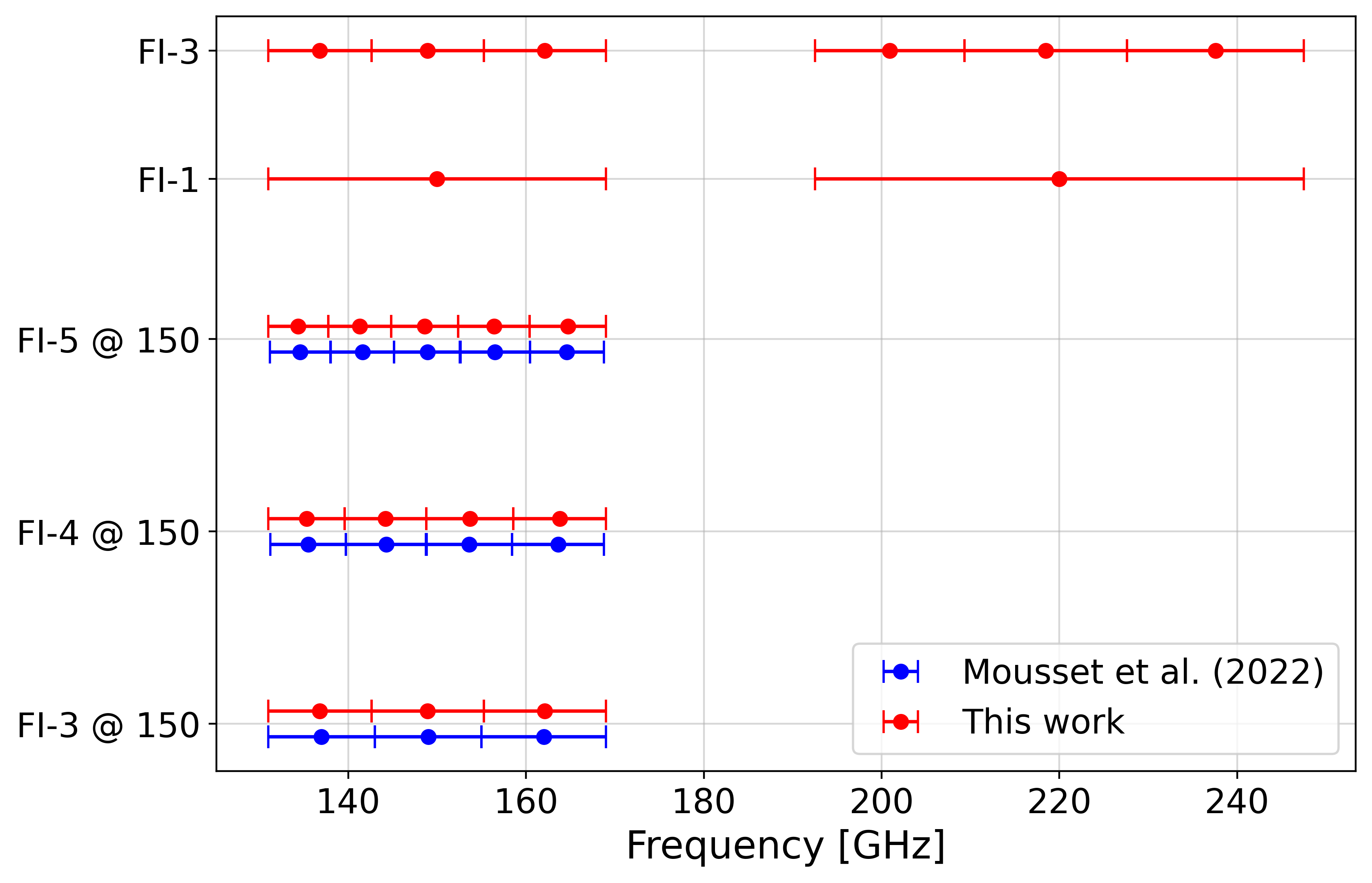}
    \caption{Comparison between sub-bands central frequencies and widths from the QUBIC official pipeline~\cite{ac}, in blue, and the ones obtained with the partitioning method proposed in this work, in red.}
    \label{Fig8}
\end{figure}

In figure~\ref{Fig8} we show a comparison between the computed sub-bands and the ones given by~\cite{ac} for the 150 GHz channel from simulated FI-3, FI-4 and FI-5 configurations. FI-4 and FI-5 configurations have not been used in the actual analysis, but are shown here as a comparison between the adopted partitioning procedure and the one used in the QUBIC official pipeline. 

In this work, the TD-3 and FI-3 configurations share the same central frequencies and bandwidths for the three sub-bands in the 150 GHz, as the hardware differences between the two instruments have no particular effect on these parameters.

\subsubsection{Noise Covariance Modelling}\label{sub4.2.3}

In a bolometric interferometer, sub-bands are reconstructed simultaneously from the same data measured in a common wide band. The way noise fluctuations are distributed into different contributions during band partitioning is strictly dependent on the constraint on the total power measured by the detectors. In practice, if a noise contribution weighs more on a sub-band, it automatically has lower impact on the neighbouring sub-bands when fitting the same data. This brings natural anti-correlations in the noise structure, which were shown by means of end-to-end simulations in~\cite{ac}.

From the same paper, we considered the correlation matrix computed from 40 noise realizations at map level using the 150 GHz channel of the FI-3 configuration. The given correlation levels between Stokes parameters and frequency sub-bands refer to a single pixel $(\mathbf{C}_\mathrm{pix})$ and to the average over all pixels in the QUBIC sky patch $(\mathbf{C}_\mathrm{avg})$. In particular, in the paper, the pixels were treated separately.

In this work, we aimed to simulate a sky patch in which the pixels exhibit different correlation levels, while remaining consistent with the representative values reported in the reference paper. Rather than performing a full set of end-to-end simulations to characterize the correlations on every pixel, we adopted a simpler and computationally cheaper procedure. For each pixel, the covariance matrix was generated as the sum of the averaged covariance $(\mathbf{C}_\mathrm{avg})$ and a random perturbation, as described below. We defined the perturbation element as
\begin{equation}\label{eq22}
\delta \mathbf{C} = |\mathbf{C}_\mathrm{pix} - \mathbf{C}_\mathrm{avg}|/4\,. 
\end{equation}
Here, the factor 4 is not physically motivated, but was chosen empirically as the smallest numerical scaling needed to get a full covariance that is positive definite given the levels of correlations and the number of pixels involved. We verified that using a larger divisor, equal to 8, does not significantly change the component separation results. Since the correlations between Stokes parameters for a single pixel shown by~\cite{ac} average out to zero over the sky patch, this scaling leads to modelling a smaller intensity to polarization leakage than the realistic case. This work, however, is not focused on Stokes correlations, but aims to study the impact of frequency correlations. On the contrary, frequency correlation levels are not significantly affected by averaging over different pixels. Thus, decreasing the absolute difference in eq.~\ref{eq22} still allows us to recover representative values on the individual pixels. 
From the perturbation element, we computed the correlation matrix for a single pixel as 
\begin{equation}\label{eq23}
\hat{\mathbf{C}}_\mathrm{pix}=\mathbf{C}_\mathrm{avg}+\delta\hat{\mathbf{C}}\,,
\end{equation}
where  $\delta\hat{\mathbf{C}}_{ij}\sim\mathcal{N}\left(0,(\delta \mathbf{C}_{ij})^2\right),\, \forall \,i,j$ entry index.
The $\hat{\mathbf{C}}_\mathrm{pix}$ matrix was then symmetrized by copying the upper triangle into the lower one, and positive definiteness was checked.

This procedure ensured that the resulting correlation matrices had realistic levels of frequency correlations for all the pixels in the sky patch, while simultaneously satisfying the constraint given by averaged correlation $\mathbf{C}_\mathrm{avg}$.
Note that we do not aim at comparing the reconstruction quality between different pixels.
Instead, each pixel provides an independent comparison between NCM assumptions for the same local sky signal and noise realization.
The total correlation matrix for the whole sky patch was then constructed by assembling together all the single-pixel perturbed matrices and making sure that positive definiteness was maintained. 
If not, a new set of single pixel correlation matrices is computed. 
Each of these was a $3N_\nu \times 3N_\nu$ matrix, making the perturbative generation of a representative full NCM computationally inexpensive compared to the multiple end-to-end simulations that would be required for a realistic characterization.

Sensitivities were computed from technical specifications and formulas given by~\cite{i,ae,af}, assuming an observed fraction of the sky $f_\mathrm{sky}=0.01$, and an observing time of $t_\mathrm{obs}^\mathrm{TD}=1$ yr or $t_\mathrm{obs}^\mathrm{FI}=3$ yr, depending on the instrumental configuration. When band partitioning is performed on a wide channel $\nu_i$, the sensitivity of the individual sub-bands $\nu_{i,j}$ grows proportionally to the number of sub-bands $N_{i,\mathrm{sub}}$ the channel has been divided into~\cite{ad}, following the equation
\begin{equation}\label{eq24}
\sigma_{i,j} = \sigma_i \cdot  \sqrt{N_{i,\mathrm{sub}}}\cdot \epsilon\left(\nu_j,N_{i,\mathrm{sub}}\right)\,.
\end{equation}
The term $\epsilon$~\cite{ac} accounts for the sub-optimality of band partitioning. In fact, part of the information is encoded in the main peak of the synthesized beam, whose position is the same for all frequencies. This contribution therefore cannot be used to distinguish the sub-bands efficiently, leading to a loss of signal in the spectral-imaging. For a 3 sub-bands partition, simulations show a 40\% degradation at 150 GHz and around 15\% at 220 GHz. However, the impact of this contribution at map level depends on the pixel position and the beam shape, in a non-trivial way. Neglecting the sub-optimality effect in the simulations slightly changes the sensitivity levels, but is not expected to significantly affect the general results of this work, which are rather related to the correlated structure of the noise.
To assess the impact of this approximation, we repeated the QUBIC-based analysis using a frequency-dependent degradation of the noise RMS, obtained by linearly interpolating representative sub-optimality factors between 150 and 220 GHz. This changes the relative sensitivity of the reconstructed sub-bands, while leaving the adopted correlation structure unchanged. We find that the qualitative behaviour of the results is not significantly affected.
Therefore, we set this contribution to $\epsilon=1$.

The sensitivities for the three simulated instrumental configurations are shown in table~\ref{tab5}.

\subsection{Component Separation Including Frequency Correlations}\label{sub4.3}

\begin{figure}[t]
    \centering
    \includegraphics[width=\linewidth]{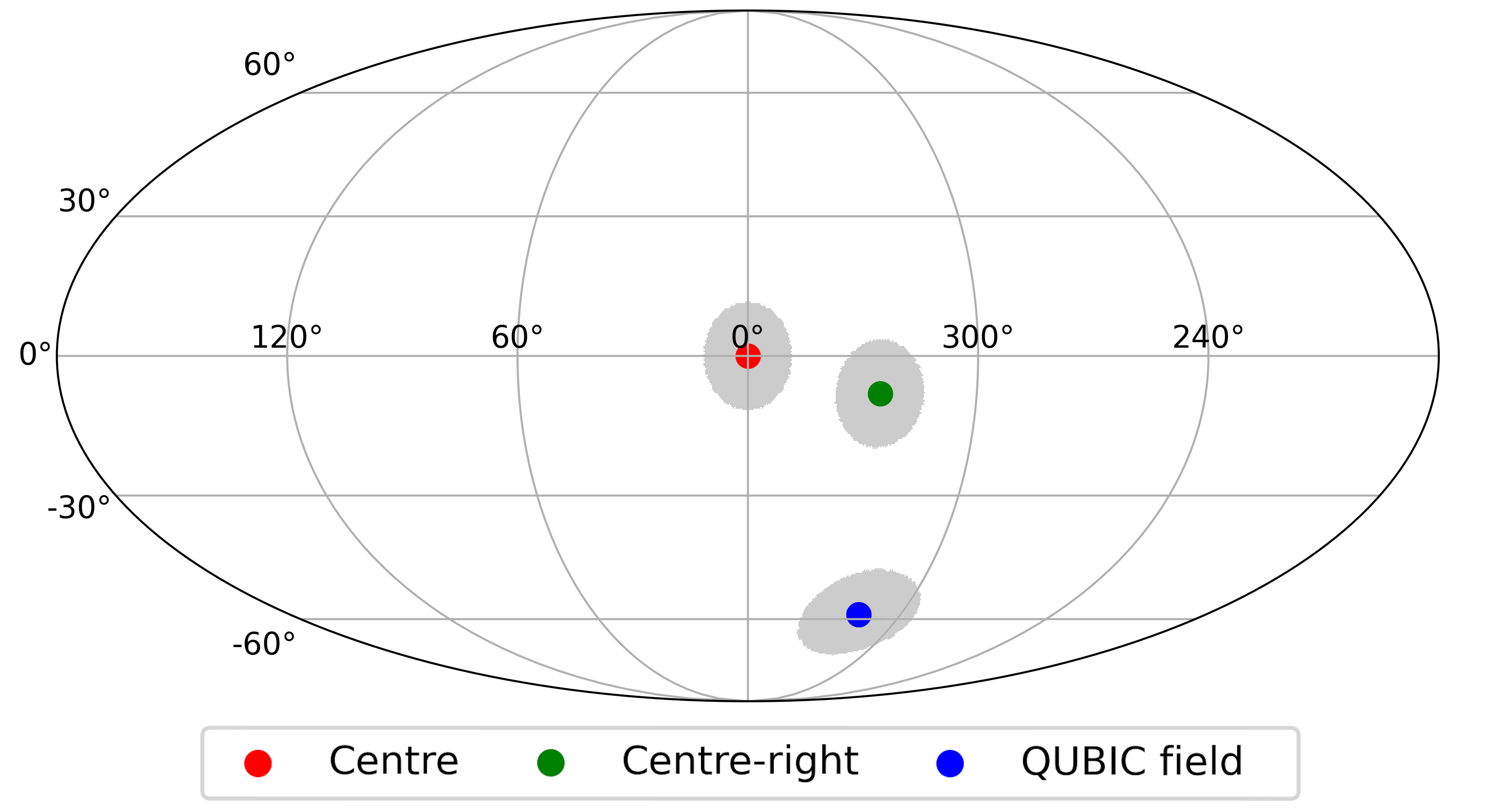}
    \caption{The three sky patches used to simulate data for the QUBIC runs.}
    \label{Fig9}
\end{figure}

As explained in the previous sections, we simulated the three instrumental configurations, namely TD-3, FI-1, and FI-3, based on the details shown in table~\ref{tab5}. 
We applied the $\nu$-generalized version of the \texttt{Commander2} code in a series of runs, where we studied the results obtained with different instrumental configurations, map resolutions, and correlation structures. The results should be interpreted as a demonstration of the impact of frequency correlations within a controlled setup, rather than a quantitative forecast of the QUBIC performance.

We chose three different sky patches to simulate the observations (figure~\ref{Fig9}), all covering 1\% of the sky. The first two patches have been defined after the ``centre'' and ``centre-right'' sky patches described by~\cite{af}, and are respectively centred at $\ell=360^\circ$, $b=0^\circ$ and $\ell=325.33^\circ$, $b=-8.02^\circ$ in Galactic coordinates. In particular, these two patches come from a list of potential targets for the QUBIC-FI observational campaign, chosen based on sky spectral variability and observational windows considerations. The third patch is the QUBIC CMB field, centred at $\ell=316.45^\circ$, $b=-58.76^\circ$. 

For simulations performed on the QUBIC patch, the same astrophysical components and models are used as for the validation runs, but using different CMB realizations. In particular, every QUBIC run was performed at $N_\mathrm{side}=64$ with the same CMB realization, in order to ensure that every difference in the results is attributable to an instrumental origin. Only the last simulation (section~\ref{sub4.3.2}) was performed at $N_\mathrm{side}=128$ using a different CMB realization.
Simulations performed on Galactic patches assumed a dust-dominated signal, therefore the CMB component has been neglected, and the analysis was performed on 3 amplitudes only.
Atmospheric contribution has not been included in the simulations.
As in the validation analysis, only the component amplitudes are sampled, while the non-linear parameters are kept fixed to their input values.
The results should therefore be interpreted as a controlled test of the impact of the covariance modelling on amplitude reconstruction, rather than as a full foreground-separation forecast for QUBIC.
The analysis on the Galactic patches should be interpreted as a simplified test in a high signal-to-noise regime, complementary to the QUBIC field. In the latter, both CMB and dust amplitudes are fitted, and the reconstruction is more affected by component degeneracies and by limited separation power provided by the available frequency channels. 
No beam smoothing was applied to the sky signal, except for the last run, discussed in section~\ref{sub4.3.2}.
Noise realizations were drawn by means of a total covariance matrix containing correlations across frequencies and Stokes parameters, defined in section~\ref{sub4.2.3}. As for the validation runs, pixels were considered independent.

Below, we list the results from different instrumental configurations. For all simulations, we ran sampling chains of 1000 iterations length, considering a 300 iterations burn-in period.

\paragraph{TD-3}\label{par4.3.1.1}
For the QUBIC Technological Demonstrator simulations, we only considered patches on the Galactic plane, since the limited sensitivities of the sub-bands make it 
unlikely to obtain a sufficient signal-to-noise ratio to perform any significant component separation on sky regions clean of galactic foregrounds.
Also, a proper parameter estimation is difficult due to the reduced frequency coverage. Therefore, this instrumental configuration shows no significant difference in the results using different NCM approximations, except for a shrinkage of all sample distributions around the analytical MLEs. This is most probably due to the sign of correlations in the instrumental noise from nearby sub-bands, as discussed in section~\ref{sub4.3.1}.

\begin{figure}[t]
    \centering
    \includegraphics[width=\linewidth]{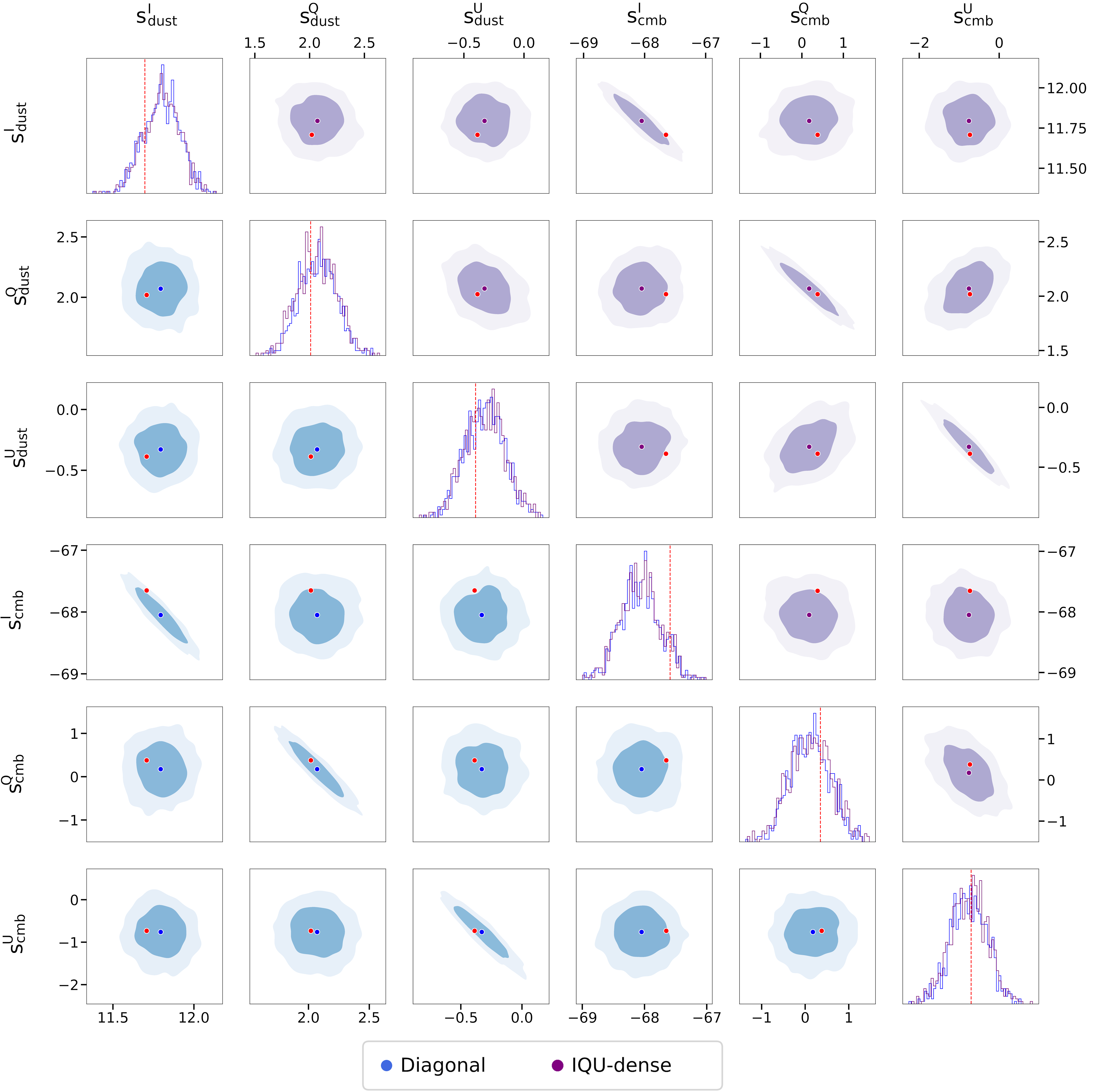}
    \caption{Comparison of joint and marginal distributions for a single noise realization and pixel from FI-1 simulations. This plot mirrors the ones shown in section~\ref{sub3.2}. Joint posterior distributions for the diagonal and IQU-dense NCM configurations (bottom-left and top-right corner, respectively in
blue and purple). Shaded contours enclose the central 68.2\% and 95.4\% intervals, respectively, going from the dark region to the light one. Coloured dots represent the corresponding analytical MLEs, while red dots and vertical lines indicate the true input parameter values used in the simulations.}
    \label{Fig10}
\end{figure}

\paragraph{FI-1}\label{par4.3.1.2}
For both QUBIC Full Instrument simulations, we performed the analysis on the ``centre-right'' Galactic sky patch and QUBIC field. The FI-1 set of runs was performed with an un-partitioned configuration, therefore no frequency correlations have been introduced. Comparing results from these simulations with the ones from the FI-3 configuration allows us to directly assess the impact of band partitioning on the estimated parameters.
In figure~\ref{Fig10} we show the joint posterior distributions of the parameters for the only two considered NCM configurations i.e. the diagonal and the IQU-dense ones.

\begin{figure}[t]
    \centering
    \includegraphics[width=\linewidth]{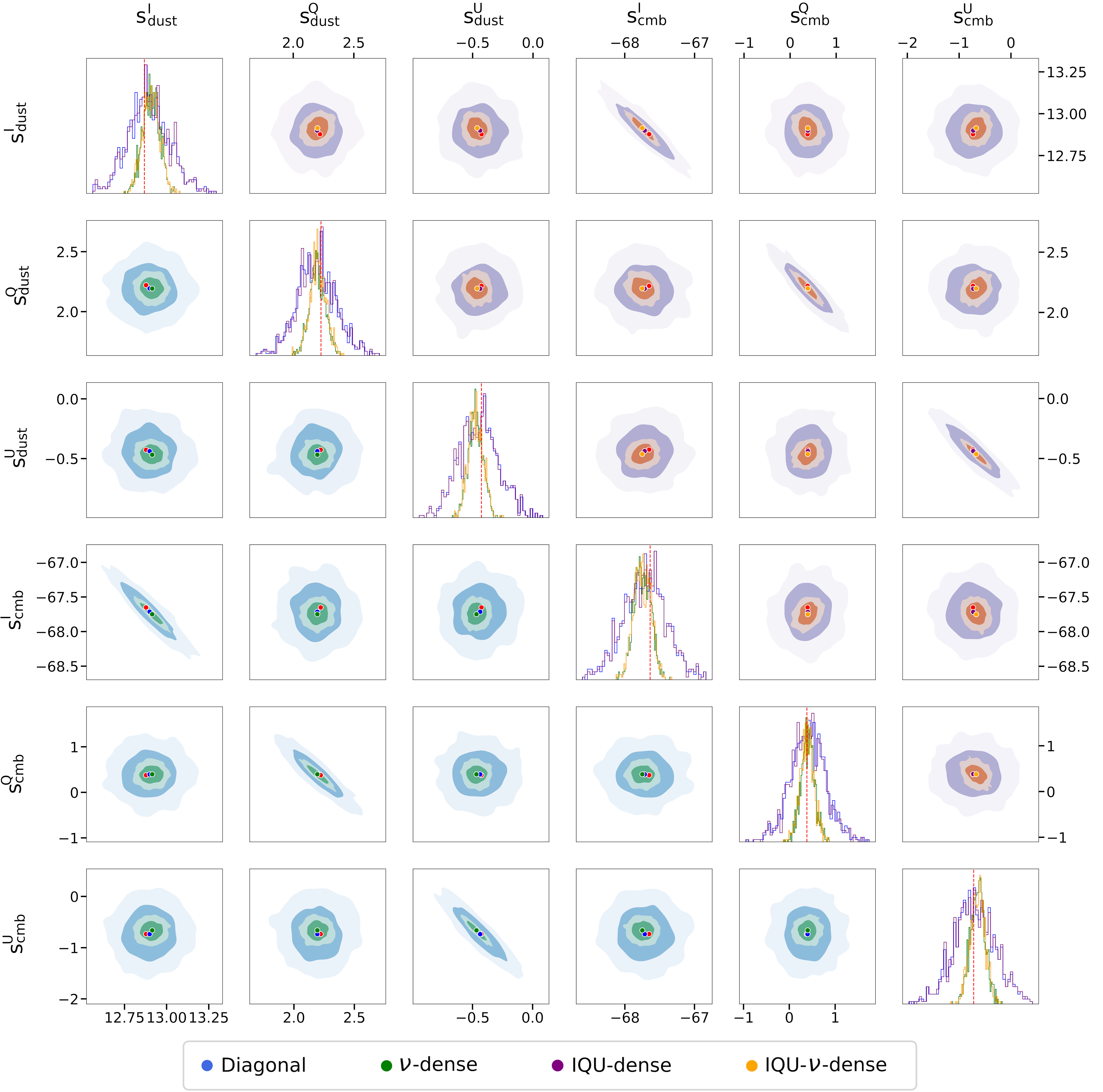}
    \caption{Comparison of joint and marginal distributions for a single noise realization and pixel for FI-3 simulations. This plot mirrors the one shown in figure~\ref{Fig10}. The distributions are shown for the diagonal and IQU-dense NCM configurations (bottom-left and top-right corner, respectively in
blue and purple), along with their $\nu$-generalized counterparts (green and orange). Shaded contours enclose the central 68.2\% and 95.4\% intervals, respectively, going from the dark region to the light one. Coloured dots represent the corresponding analytical MLEs, while red dots and vertical lines indicate the true input parameter values used in the simulations.}
    \label{Fig11}
\end{figure}

\paragraph{FI-3}\label{par4.3.1.3}
The FI-3 simulations were performed on the same sky patches as the FI-1 configurations. In this case, FI wide bands were partitioned in 3 sub-bands each, so results were also obtained for the $\nu$-generalized cases. figure~\ref{Fig11} shows the joint posterior distributions for all parameters and NCM configurations, computed on the same pixel considered for the plotted results in figure~\ref{Fig10}. All distributions appear more centred around true values with respect to FI-1, as a direct effect of spectral imaging. Indeed, partitioning into sub-bands is proven to help break degeneracies in component separation despite increasing the overall noise level~\cite{ad}.

As for TD-3 simulations, $\nu$-generalized distributions are narrower than legacy ones. In table~\ref{tab7} we gathered results from the three instrumental configurations. The values refer to the MLE residual $-$ as defined in eq.~\ref{eq15} $-$ on the central pixel of the corresponding sky patch. Superscripts and subscripts indicate the 68th percentiles of the left and right tails of the sample distributions.

\begin{figure}[t]
    \centering
    \includegraphics[width=0.73\linewidth]{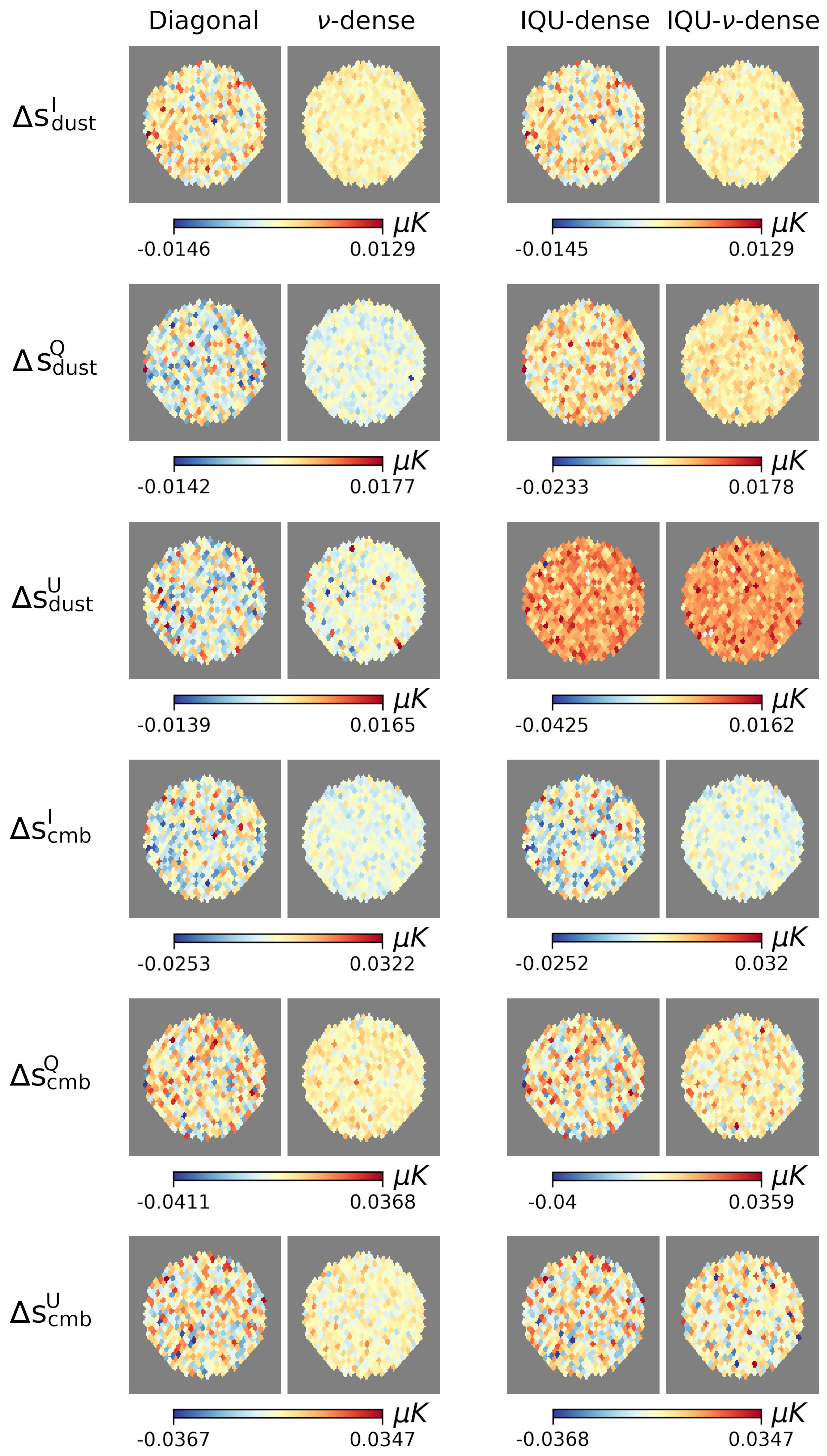}
    \caption{Maps of the MLE residuals for the FI-3 run on the QUBIC patch. Temperature scales for each legacy NCM configuration and its $\nu$-generalized counterpart are kept the same to highlight the comparison.}
    \label{Fig12}
\end{figure}

In addition to systematically reducing the overall uncertainties, $\nu$-generalized analyses actually produced chain means which are closer to the analytical MLE for most of the pixels, leading to cleaner MLE residual maps (figure~\ref{Fig12}).

\begin{table}[t]
\hspace{-29pt}
\renewcommand{\arraystretch}{1.5}
\begin{tabular}{|cccccccccc}
\cline{3-8}
\multicolumn{1}{}{}   &      & \multicolumn{2}{|c}{TD-3} &   \multicolumn{2}{|c}{FI-1} & \multicolumn{2}{|c|}{FI-3} \\ \hline
\multicolumn{1}{|c|}{$\mathbf{\Delta s} \,\,(\boldsymbol{\mu}\mathrm{\textbf{K}})$} & \multicolumn{1}{c|}{\textbf{NCM conf.}}  &     \multicolumn{1}{|c}{centre}      & \multicolumn{1}{c|}{centre-right} & \multicolumn{1}{c}{centre-right}   & \multicolumn{1}{c|}{QUBIC}   &\multicolumn{1}{c}{centre-right} & \multicolumn{1}{c|}{QUBIC} \\ \hline
\multirow{4}{*}{$\mathrm{\Delta s}_\mathrm{dust}^\mathrm{I}$} & \multicolumn{1}{|c|}{Diagonal} & $0^{+46}_{-43}$ & \multicolumn{1}{c|}{$0^{+47}_{-46}$}& 
$0^{+0.04}_{-0.04}$
& \multicolumn{1}{c|}{
$0^{+0.1}_{-0.1}$
} & 
$0^{+0.04}_{-0.04}$
& 
\multicolumn{1}{c|}{$0.01^{+0.10}_{-0.10}$}

\\

  & \multicolumn{1}{|c|}{$\nu$-dense} & 
$0^{+15}_{-15}$
& \multicolumn{1}{c|}{
$1^{+14}_{-14}$
}
 & $-$ & \multicolumn{1}{c|}{$-$} &
$0^{+0.02}_{-0.02}$
  &
\multicolumn{1}{c|}{$0^{+0.04}_{-0.04}$}
  \\
  
  & \multicolumn{1}{|c|}{IQU-dense} & 
  $-1^{+45}_{-43}$
 & \multicolumn{1}{c|}{
   $0^{+48}_{-46}$
} & 
   $0^{+0.04}_{-0.04}$
 & \multicolumn{1}{c|}{
    $0^{+0.1}_{-0.1}$
} &
  $0^{+0.04}_{-0.04}$
&
    \multicolumn{1}{c|}{$0^{+0.10}_{-0.10}$}
 \\

  & \multicolumn{1}{|c|}{IQU-$\nu$-dense} & 
    $0^{+15}_{-15}$  
   & \multicolumn{1}{c|}{
      $1^{+14}_{-14}$}
    & $-$ & \multicolumn{1}{c|}{$-$} & 
    $0^{+0.02}_{-0.02}$
& 
    \multicolumn{1}{c|}{$0.01^{+0.04}_{-0.04}$}
\\ \hline
  

\multirow{4}{*}{$\mathrm{\Delta s}_\mathrm{dust}^\mathrm{Q}$} & \multicolumn{1}{|c|}{Diagonal} & 
    $4^{+63}_{-61}$ 
 & \multicolumn{1}{c|}{
    $2^{+65}_{-65}$ 
  }&
     $0^{+0.05}_{-0.05}$
 & \multicolumn{1}{c|}{
    $-0.1^{+0.1}_{-0.1}$ 
} & 
     $0.01^{+0.06}_{-0.06}$
& 
     \multicolumn{1}{c|}{$0^{+0.13}_{-0.13}$}
\\

  & \multicolumn{1}{|c|}{$\nu$-dense} & 
     $0^{+18}_{-19}$  
 & \multicolumn{1}{c|}{
      $1^{+23}_{-24}$
} & $-$ & \multicolumn{1}{c|}{$-$} & 
      $0^{+0.02}_{-0.02}$ 
& 
     \multicolumn{1}{c|}{$0^{+0.06}_{-0.06}$}
\\
  
  & \multicolumn{1}{|c|}{IQU-dense} & 
     $3^{+63}_{-61}$  
& \multicolumn{1}{c|}{
      $1^{+65}_{-65}$  
} & 
      $0^{+0.05}_{-0.05}$
 & \multicolumn{1}{c|}{
      $-0.1^{+0.1}_{-0.1}$
} & 
     $0.01^{+0.06}_{-0.06}$
& 
     \multicolumn{1}{c|}{$0^{+0.14}_{-0.13}$}
\\

  & \multicolumn{1}{|c|}{IQU-$\nu$-dense} & 
       $0^{+20}_{-19}$ 
& \multicolumn{1}{c|}{
      $1^{+23}_{-24}$
} & $-$ & \multicolumn{1}{c|}{$-$} & 
      $0^{+0.02}_{-0.02}$
& 
     \multicolumn{1}{c|}{$0^{+0.06}_{-0.06}$}
\\ \hline
  

\multirow{4}{*}{$\mathrm{\Delta s}_\mathrm{dust}^\mathrm{U}$} & \multicolumn{1}{|c|}{Diagonal} & 
      $2^{+61}_{-61}$
& \multicolumn{1}{c|}{
      $-2^{+63}_{-66}$
} & 
      $0^{+0.05}_{-0.05}$
& \multicolumn{1}{c|}{
      $0^{+0.1}_{-0.1}$
} & 
        $0.01^{+0.06}_{-0.05}$
& 
      \multicolumn{1}{c|}{$0^{+0.13}_{-0.14}$}
\\

  & \multicolumn{1}{|c|}{$\nu$-dense} & 
      $0^{+18}_{-16}$  
& \multicolumn{1}{c|}{
      $0^{+19}_{-19}$
} & $-$ & \multicolumn{1}{c|}{$-$} & 
      $0^{+0.02}_{-0.02}$  
& 
      \multicolumn{1}{c|}{$0^{+0.06}_{-0.06}$}
\\
  
  & \multicolumn{1}{|c|}{IQU-dense} & 
      $3^{+61}_{-61}$  
 & \multicolumn{1}{c|}{
      $-3^{+63}_{-64}$  
} & 
      $0^{+0.05}_{-0.05}$ 
& \multicolumn{1}{c|}{
      $0^{+0.1}_{-0.1}$ 
} & 
       $0^{+0.06}_{-0.06}$
& 
      \multicolumn{1}{c|}{$0.01^{+0.13}_{-0.14}$}
\\

  & \multicolumn{1}{|c|}{IQU-$\nu$-dense} &
      $1^{+17}_{-16}$  
& \multicolumn{1}{c|}{
      $1^{+19}_{-19}$
} & $-$ & \multicolumn{1}{c|}{$-$} & 
      $0.01^{+0.02}_{-0.02}$
& 
      \multicolumn{1}{c|}{$0^{+0.06}_{-0.06}$}
\\ \hline
  

\multirow{4}{*}{$\mathrm{\Delta s}_\mathrm{cmb}^\mathrm{I}$} & \multicolumn{1}{|c|}{Diagonal} & $-$ & \multicolumn{1}{c|}{$-$} & $-$ & \multicolumn{1}{c|}{
      $0^{+0.3}_{-0.3}$
} & $-$ & 
      \multicolumn{1}{c|}{$-0.01^{+0.25}_{-0.23}$}
      \\

  & \multicolumn{1}{|c|}{$\nu$-dense} & $-$ & \multicolumn{1}{c|}{$-$} & $-$ & \multicolumn{1}{c|}{$-$} & $-$ & 
        \multicolumn{1}{c|}{$0^{+0.08}_{-0.08}$}
  \\
  
  & \multicolumn{1}{|c|}{IQU-dense} & $-$ & \multicolumn{1}{c|}{$-$} & $-$ & \multicolumn{1}{c|}{
      $0^{+0.3}_{-0.3}$  
} & $-$ & 
      \multicolumn{1}{c|}{$0^{+0.25}_{-0.23}$}
\\

  & \multicolumn{1}{|c|}{IQU-$\nu$-dense} & $-$ & \multicolumn{1}{c|}{$-$} & $-$ & \multicolumn{1}{c|}{$-$} & $-$ & 
        \multicolumn{1}{c|}{$-0.01^{+0.09}_{-0.08}$}
  \\ \hline
  

\multirow{4}{*}{$\mathrm{\Delta s}_\mathrm{cmb}^\mathrm{Q}$} & \multicolumn{1}{|c|}{Diagonal} & $-$ & \multicolumn{1}{c|}{$-$} & $-$ & \multicolumn{1}{c|}{
      $0^{+0.3}_{-0.4}$
} & $-$ & 
      \multicolumn{1}{c|}{$0^{+0.3}_{-0.3}$}
\\

  & \multicolumn{1}{|c|}{$\nu$-dense} & $-$ & \multicolumn{1}{c|}{$-$} & $-$ & \multicolumn{1}{c|}{$-$} & $-$ & 
        \multicolumn{1}{c|}{$0^{+0.1}_{-0.1}$}
  \\
  
  & \multicolumn{1}{|c|}{IQU-dense} & $-$ & \multicolumn{1}{c|}{$-$} & $-$ & \multicolumn{1}{c|}{
        $0^{+0.4}_{-0.4}$
 } & $-$ & 
       \multicolumn{1}{c|}{$0^{+0.3}_{-0.3}$}
 \\

  & \multicolumn{1}{|c|}{IQU-$\nu$-dense} & $-$ & \multicolumn{1}{c|}{$-$} & $-$ & \multicolumn{1}{c|}{$-$} & $-$ & 
        \multicolumn{1}{c|}{$0^{+0.1}_{-0.2}$}
  \\ \hline
  

\multirow{4}{*}{$\mathrm{\Delta s}_\mathrm{cmb}^\mathrm{U}$} & \multicolumn{1}{|c|}{Diagonal} & $-$ & \multicolumn{1}{c|}{$-$} & $-$ & \multicolumn{1}{c|}{
      $0^{+0.4}_{-0.4}$
} & $-$ & 
      \multicolumn{1}{c|}{$0^{+0.4}_{-0.3}$}
\\

  & \multicolumn{1}{|c|}{$\nu$-dense} & $-$ & \multicolumn{1}{c|}{$-$} & $-$ & \multicolumn{1}{c|}{$-$} & $-$ & 
        \multicolumn{1}{c|}{$0^{+0.1}_{-0.1}$}
  \\
  
  & \multicolumn{1}{|c|}{IQU-dense} & $-$ & \multicolumn{1}{c|}{$-$} & $-$ & \multicolumn{1}{c|}{
        $0^{+0.4}_{-0.4}$
} & $-$ & 
        \multicolumn{1}{c|}{$-0.1^{+0.4}_{-0.3}$}
\\

  & \multicolumn{1}{|c|}{IQU-$\nu$-dense} & $-$ & \multicolumn{1}{c|}{$-$} & $-$ & \multicolumn{1}{c|}{$-$} & $-$ & 
        \multicolumn{1}{c|}{$0^{+0.1}_{-0.2}$}
  \\ \hline
  

\end{tabular}
\caption{MLE residuals on each sky patch central pixel, for every instrumental and NCM configuration. Errors refer to left and right 68th percentiles.} 
\label{tab7}
\end{table}

\subsubsection{Positive Correlation}\label{sub4.3.1}
In order to investigate the reason for the narrower distributions in the $\nu$-generalized runs, we simulated a set of frequency maps for which the sign of the noise correlations had been flipped compared to the FI-3 case, i.e. with positive rather than negative cross-band noise correlations. All other simulation details are the same as for the FI-3 runs. Note that this configuration is not physical, since for QUBIC it has been confirmed by laboratory tests that band partitioning introduces strong anti-correlated noise in the sub-bands. This test has been conducted on the QUBIC patch only. 

While the anti-correlated FI-3 distributions were reduced in width by around 60\% , we found that $\nu$-generalized distributions from this positive-correlated analysis became around 40\% wider than their legacy counterparts.
We recall that in section~\ref{Sec3} distribution widths during the validation phases remained almost unchanged despite a 30\% positive correlation between frequency channels. 
This behaviour can be qualitatively understood from a simple two-channel example. 
For two maps with noise variances $\sigma_1^2$ and $\sigma_2^2$, and correlation coefficient $\rho$, the covariance matrix is
\begin{equation}\label{eq25}
\mathbf{N} =
\begin{pmatrix}
\sigma_1^2 & \rho \sigma_1\sigma_2 \\
\rho \sigma_1\sigma_2 & \sigma_2^2
\end{pmatrix}\,,
\end{equation}
and its inverse is
\begin{equation}\label{eq26}
\mathbf{N}^{-1}
=
\frac{1}{1-\rho^2}
\begin{pmatrix}
\sigma_1^{-2} & -\rho(\sigma_1\sigma_2)^{-1} \\
-\rho(\sigma_1\sigma_2)^{-1} & \sigma_2^{-2}
\end{pmatrix}\,.
\end{equation}
The contribution to the likelihood is therefore
\begin{equation}\label{eq27}
\chi^2 =
\frac{1}{1-\rho^2}
\left[
\frac{r_1^2}{\sigma_1^2}
+
\frac{r_2^2}{\sigma_2^2}
-
2\rho\frac{r_1r_2}{\sigma_1\sigma_2}
\right]\,,
\end{equation}
where $r_i$ is the residual in channel $i$. 
A positive (or negative) sign for $\rho$ determines whether residuals with the same (or opposite) signs in the two channels are penalized more.
This shows why changing the sign of cross-channel correlations can modify the posterior shape. 

However, the CMB-S4-like validations and the QUBIC-based sign-flip test show that the direct impact of NCM approximation on joint posterior distributions is not determined by the sign of the correlations alone, but depends on each individual covariance structure and must be assessed for each specific instrumental configuration.

\subsubsection{High Resolution Smoothed Maps}\label{sub4.3.2}
For the final set of analyses, we simulated frequency maps at $N_\mathrm{side}=128$ using the same configuration as the anti-correlated FI-3 case, but also including beam smoothing of the sky signal. These runs were performed to test the handling of beam convolution and deconvolution within the new \texttt{Commander2} implementation under more realistic observing conditions. Since the noise covariance matrix remained uncorrelated between pixels and only the sky signal was smoothed, component separation was still performed independently at each pixel. Therefore, this higher resolution analysis also aims to study differences in parameter estimation with a lower signal-to-noise ratio.
During partitioning, the angular resolution of the sub-bands slightly varies with frequency~\cite{ac}, but still remaining close to the value assumed at the wide band central frequency. Thus, assuming a Gaussian beam, the FWHM for the low-frequency sub-bands was around 23 arcmin. This is comparable to the characteristic pixel size at $N_\mathrm{side}=128$, which is 27.48 arcmin, leading to ringing and numerical instabilities. To mitigate these effects, the sky signal was smoothed with a Gaussian beam of 68.71 arcmin in FWHM, corresponding to 2.5 times the HEALPix pixel size. Spherical harmonic transforms were performed up to $\ell_\mathrm{max}=2.5\times N_\mathrm{side}=320$.

\begin{figure}[!t]
    \centering
    \includegraphics[width=\linewidth]{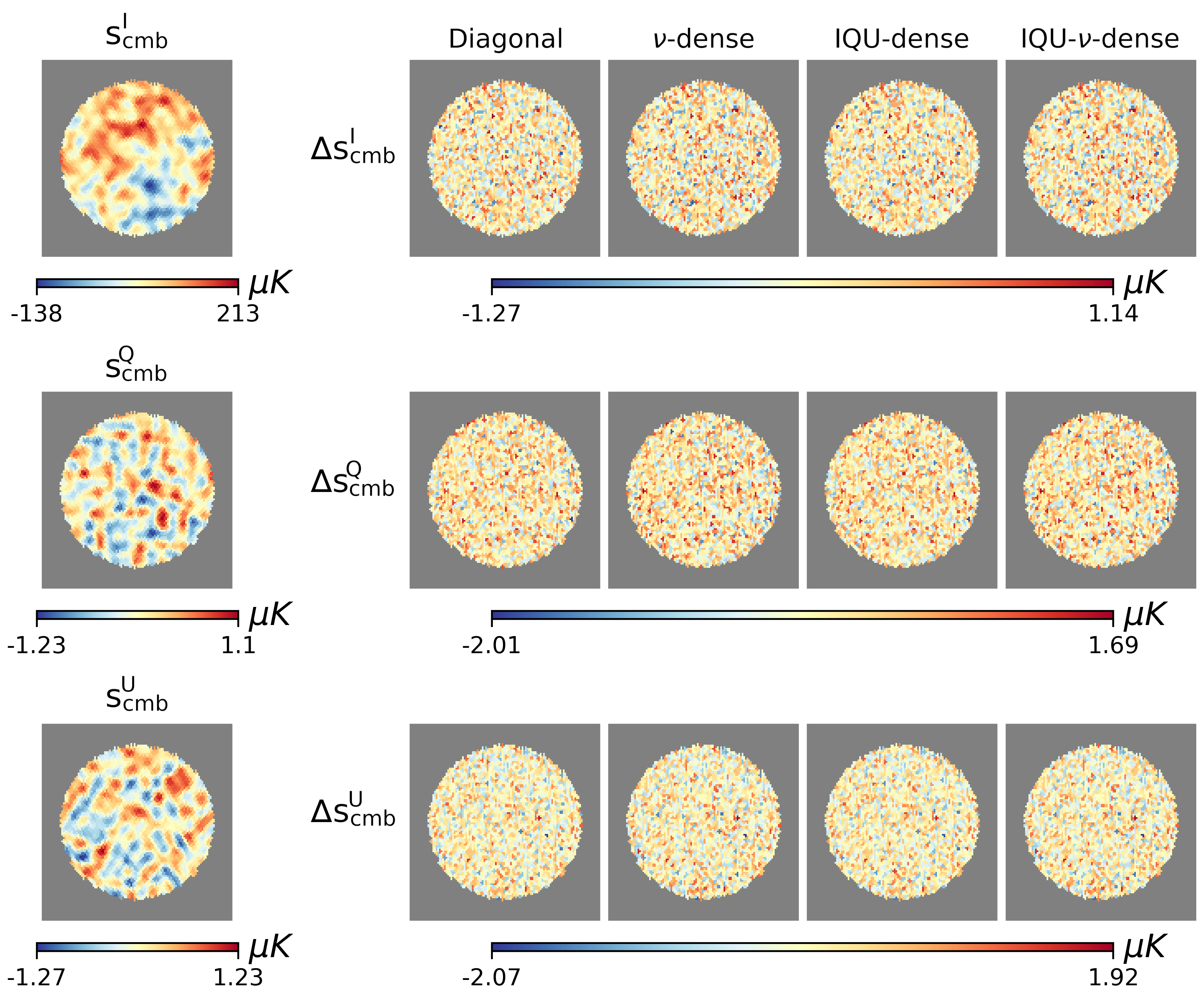}
    \caption{Maps of the differences between chain means and input values for the CMB amplitudes from the high-resolution FI-3 runs (on the right). Rows separate maps by Stokes parameter, and display results on the QUBIC sky patch. The first column on the left shows the true input smoothed CMB maps, in order to compare the amplitude of the residuals with the input signal level after the component separation.}
    \label{Fig13}
\end{figure}

In figure~\ref{Fig13} we show the differences between chain means and true input values for the smoothed CMB amplitudes, compared with the true input smoothed CMB intensity map, for all NCM configurations. In table~\ref{tab8} we list the average residuals computed over the whole patch. 
In this case, we find no significant improvement in the resulting intensity maps when frequency correlations are accounted for.
Polarization amplitudes are poorly reconstructed, and the associated maps appear to be noise dominated for scales smaller than 2 degrees.


\section{Conclusions}\label{Sec5}

This work presents an extension of the \texttt{Commander2} component separation framework to support instrumental noise covariance matrices including correlations between frequency channels. While most current parametric component separation analyses assume noise to be independent across frequencies, this approximation may not be sufficient for future experiments, as the required level of accuracy is increasing. This is particularly relevant for instruments employing interferometric or multi-band bolometric technologies, for which cross-channel correlations can naturally arise.

The generalized implementation introduced here removes the implicit assumption of likelihood separability across frequency channels by allowing for a noise covariance structure that is fully dense in pixels, Stokes parameters, and frequencies. This required modifications to the parallel architecture and memory handling of \texttt{Commander2}, enabling the consistent propagation of correlated noise information into parameter estimation.

We validated the code through a controlled set of low-resolution simulations based on a simplified CMB-S4-like configuration. Convergence diagnostics show excellent agreement between chains, with Gelman–Rubin statistics remaining close to unity for all sampled parameters. We further assessed the correctness of the sampling by comparing MCMC estimates with analytical maximum-likelihood solutions. The results show that the generalized implementation recovers the expected posterior distributions within the statistical limits of the adopted setup.
In addition, analytical estimates generally show more compact distributions around the true input values when frequency correlations are included.

\begin{table}[!t]
\centering
  \renewcommand{\arraystretch}{1.2}
  \begin{tabular}{cccc|}
   \cline{2-4}
    & \multicolumn{3}{|c|}{Average relative residual (nK)} \\
    \multicolumn{1}{c|}{} & $<\Delta \mathrm{s}_\mathrm{cmb}^\mathrm{I}>$ & $<\Delta \mathrm{s}_\mathrm{cmb}^\mathrm{Q}>$ & $<\Delta\mathrm{s}_\mathrm{cmb}^\mathrm{U}>$ \\ 
    
    \hline
    \multicolumn{1}{|c|}{Diagonal}        & $-1\pm353$ & $6\pm504$ & $-21\pm502$ \\ \cline{2-4}
    \multicolumn{1}{|c|}{$\nu$-dense}     & $\hspace{6pt}\,4\pm374$ & $4\pm545$ & $-21\pm532$\\ \cline{2-4}
    \multicolumn{1}{|c|}{IQU-dense}       & $-2\pm352$ & $7\pm506$ & $-20\pm507$\\ \cline{2-4}
     \multicolumn{1}{|c|}{IQU-$\nu$-dense} & $\hspace{6pt}\,4\pm374$ & $3\pm542$ & $-21\pm530$\\  \cline{1-4}
  \end{tabular}
\caption{CMB amplitude residuals from the high-resolution FI-3 runs. Results are averaged over the QUBIC sky patch.} 
\label{tab8}
\end{table}

We then applied the extended framework to instrument-motivated simulations based on QUBIC, a bolometric interferometer for which frequency-correlated noise represents a particularly relevant test case. Using simplified but representative instrumental configurations derived from the official specifications, we explored the effect of frequency correlations on component separation. The results indicate that accounting for these cross-channel correlations during component separation leads to a better estimation of the parameters uncertainties, and improves the agreement of the recovered sky amplitudes with their input values in the case of a sufficiently high signal-to-noise ratio.

As noted above, the QUBIC results are illustrative rather than a performance forecast; the key finding is that accounting for these cross-channel correlations improves both the parameter uncertainties and, at sufficiently high signal-to-noise, the recovered amplitudes themselves. A natural next step is to apply the generalized implementation to frequency maps and pixel-dependent noise covariance matrices produced by dedicated end-to-end map-making simulations.

The implementation developed in this work provides a general framework for including frequency correlations in parametric component separation analyses within the \texttt{Commander2} software. Since the effect treated here is independent of the physical origin of the correlations, the code may be relevant for a broad range of future CMB experiments. At the same time, the associated computational cost suggests that the practical benefit of including frequency correlations should be evaluated on a case-by-case basis.

\appendix
\section{Computing performance}\label{Appendix A}
The new implementation of the code has two major computational consequences, namely on the memory usage and the computing time. 

\paragraph{Memory requirements}
In the $\nu$-generalized implementation, the number of MPI processes does not need to be increased with respect to the legacy implementation: in the runs presented in this work, we assign one MPI process to each frequency band in both cases. However, while in the legacy configuration each process only needs the information associated with its own frequency channel, in the generalized case they must load the corresponding data for all channels. This include
sky maps, beam profiles, and foreground model templates for all channels. These are objects with size at most equal to $12\times N_\mathrm{side}^2$ per Stokes component, and are therefore tractable for typical use cases. 
By contrast, the peak memory requirement associated with the noise covariance blocks is not increased by the $\nu$-generalization. 
Each MPI process iteratively loads and unloads the relevant blocks $\left( \mathbf{N}^{-1}\right)_{\nu_i \nu_j}$ of the inverse noise covariance matrix at each likelihood evaluation step.
These blocks have a fixed size by design, with dimension $(N_{\mathrm{pix}} \times N_{\mathrm{Stokes}}) \times (N_{\mathrm{pix}} \times N_{\mathrm{Stokes}})$, where $N_\mathrm{pix}$ is the number of pixels in the unmasked region of the map.
This is the same size as the NCM that is loaded in the legacy IQU-dense configuration. Since only one block is kept in memory at a time, the peak memory usage for the noise covariance remains constant throughout the run. However, since the blocks must be accessed repeatedly during the analysis, this strategy introduces substantial I/O overhead.
In table~\ref{tabA1} we show a rough comparison between the relevant memory contributions.
The values assume double-precision floating point quantities, 10 frequency channels, and an observed sky fraction of 1\%. The latter affects the covariance block size, which depends on the number of unmasked pixels, but not the map-like objects, which are assumed to be stored in full-sky format.
Let $x$ be the size of a single full-sky IQU map at one frequency.
The total map-like memory per frequency is of the order of $10\,x$, accounting for the different sky maps, templates, masks, and RMS maps used in the code. Therefore, the legacy case would require approximately $10\,x$ per MPI process, while the $\nu$-generalized case would require approximately $100\,x$.

\begin{table}[!t]
\centering
  \renewcommand{\arraystretch}{1.2}
  \begin{tabular}{|c|c|c|c|c|}
   \hline
   $N_\mathrm{side}$ &  {\begin{tabular}[c]{@{}c@{}} Single \\ IQU map\end{tabular}} & {\begin{tabular}[c]{@{}c@{}} All maps \\ (legacy)\end{tabular}} & {\begin{tabular}[c]{@{}c@{}}All maps \\ ($\nu$-gen.)\end{tabular}} & NCM \\ \hline
    4 & $5\,$KB & $50\,$KB & $500\,$KB & $0.3\,$KB \\
    16 & $80\,$KB & $800\,$KB & $8\,$MB & $70\,$KB \\
    64 & $1.5\,$MB & $15\,$MB & $150\,$MB & $20\,$MB \\
    256 & $20\,$MB & $200\,$MB & $2\,$GB & $4.5\,$GB \\
    512 & $80\,$MB & $800\,$MB & $8\,$GB & $71\,$GB \\
    \hline
  \end{tabular}
\caption{Approximate memory usage per MPI process of map-like quantities and of a single covariance block. The estimates assume double-precision floating point values, 10 frequency channels, and an observed sky fraction of 1\%. The NCM memory contribution is the same for both legacy and $\nu$-generalized configurations.} 
\label{tabA1}
\end{table}

This comparison shows that the additional memory required by the $\nu$-generalized implementation is dominated by the storage of map-like quantities from all frequency channels only at resolution for which the memory usage is not itself a limiting factor. At higher resolution, or for larger sky fractions, the dominant limitation is instead set by the size of the covariance blocks. This limitation is already present in the legacy IQU-dense case, since the size of a single covariance block is unchanged by the $\nu$-generalization.

\paragraph{Computing time}
The main additional cost introduced by the new implementation is not the peak memory usage, but the increase in computing time.
Part of this increase is due to the need to iteratively load and unload the relevant blocks of the inverse noise covariance matrix at each likelihood evaluation step. 
However, the dominant contribution to the additional computing time arises from the convergence of the Conjugate Gradient (CG) solver. The more complex structure of the cross-frequency covariance blocks generally increases the number of CG iterations required to reach a fixed residual tolerance. 
The overall runtime depends strongly on the specific analysis setup, including the structure of the noise covariance, the number of frequency channels, the number of CPUs per channel, the map resolution and the processing mask, the number of sampling iterations, the presence of beam smoothing, and the CG tolerance. It is therefore not meaningful to quote a single universal factor to quantify the code performance.
All simulations were performed on the same computer node equipped with two Intel Xeon E5-2697 v4 CPUs (2.3 GHz, 18 cores each) and 128 GB of RAM. 
The analysis was parallelized by assigning one MPI process to each frequency band.

We compared different NCM configurations using identical datasets and sampling hyperparameters, in particular with a CG tolerance of $\epsilon = 10^{-6}$.
The reported values correspond to the average over all iterations of the wall-clock times, computed excluding initialization and preconditioning. 
In particular, we partially mitigated the computational cost by retaining in the code the default preconditioning strategy adopted by \texttt{Commander2}. This consists of constructing the CG solver preconditioner at each sampling iteration from the RMS map of the primary frequency channel only.

Table~\ref{tabA2} reports the average wall-clock time per sampling iteration for the validation runs, performed on a single pixel at $N_\mathrm{side}=4$ (section~\ref{Sec3}). Phase I includes frequency correlations between adjacent channels only, whereas in Phase II all frequency channels are correlated.

\begin{table}[!t]
\centering
  \renewcommand{\arraystretch}{1.2}
  \begin{tabular}{ccc|c|}
   \cline{3-4}
    &                 & \multicolumn{1}{|c|}{Phase I} & Phase II \\ \hline
    \multicolumn{1}{|c|}{\multirow{4}{*}{\begin{tabular}[|c]{@{}c@{}}Average time \\ per iter. {(}s{)}\end{tabular}}} & \multicolumn{1}{|c|}{Diagonal}        & 0.020 & 0.020 \\ \cline{2-4}
    \multicolumn{1}{|c|}{}                                                                                           & \multicolumn{1}{|c|}{$\nu$-dense}     & 1.385 & 1.370 \\ \cline{2-4}
    \multicolumn{1}{|c|}{}                                                                                           & \multicolumn{1}{|c|}{IQU-dense}       & 0.021 & 0.021 \\ \cline{2-4}
    \multicolumn{1}{|c|}{}                                                                                           & \multicolumn{1}{|c|}{IQU-$\nu$-dense} & 4.414 & 13.999 \\  \cline{1-4}
  \end{tabular}
\caption{Average computing times per iteration for the validation phases.} 
\label{tabA2}
\end{table}

The diagonal and IQU-dense configurations have nearly identical computing times in the two phases. Since these configurations do not include cross-frequency covariance blocks, their computational treatment is unchanged when moving from Phase I to Phase II. The $\nu$-generalized configurations are substantially more expensive than their legacy counterparts. This increase is primarily due to the larger number of CG iterations required to reach the adopted residual tolerance. 

Table~\ref{tabA3} reports the average timings per iteration for two representative QUBIC-based simulations, using the FI-3 instrumental setup. The first is performed at $N_\mathrm{side} = 64$ without beam smoothing, while the second is performed at $N_\mathrm{side}=128$ including beam smoothing.

\begin{table}[!t]
\centering
  \renewcommand{\arraystretch}{1.2}
  \begin{tabular}{cc|c|c|}
   \cline{3-4}
    &                 & $N_\mathrm{side}=64$ & $N_\mathrm{side}=128$ \\ 
&                 & \textit{unsmoothed} & \textit{smoothed} \\     
    
    \hline
    \multicolumn{1}{|c|}{\multirow{4}{*}{\begin{tabular}[c]{@{}c@{}}Average time \\ per iter. {(}s{)}\end{tabular}}} & Diagonal        & 0.48 & 1.61 \\ \cline{2-4}
    \multicolumn{1}{|c|}{}                                                                                           & $\nu$-dense     & 5.99 & 122.76 \\ \cline{2-4}
    \multicolumn{1}{|c|}{}                                                                                           & IQU-dense       & 0.95 & 10.00 \\ \cline{2-4}
    \multicolumn{1}{|c|}{}                                                                                           & IQU-$\nu$-dense & 6.82 & 239.68 \\  \cline{1-4}
  \end{tabular}
\caption{Average computing times per iteration for the FI-3 QUBIC runs.} 
\label{tabA3}
\end{table}

For the diagonal configuration, increasing the resolution of a factor 2 raises the average time per iteration by approximately a factor 4, as expected. For the IQU-dense, $\nu$-dense, and fully dense configurations, instead, the computation time increases by factors of approximately 10, 20, and 35, respectively. As in the validation runs, the principal cause is the larger number of CG iterations required for convergence. In addition, the second run requires the application of the beam operator. These benchmarks show that the computational overhead of the $\nu$-generalized implementation is controlled not only by the dimensions and density of the noise covariance matrix, but primarily by the conditioning of the resulting linear system. A natural direction for future development is therefore the construction of an optimized preconditioner specifically designed for frequency-correlated covariance structures.

\acknowledgments

This paper and related research have been conducted during and with the support of the Italian national inter-university PhD programme in Space Science and Technology.\\
This article was produced while attending the PhD programme in PhD in Space Science and Technology at the University of Trento, Cycle XXXIX, with the support of a scholarship financed by the Ministerial Decree no. 118 of 2nd March 2023, based on the NRRP - funded by the European Union - NextGenerationEU - Mission 4 "Education and Research", Component 1 "Enhancement of the offer of educational services: from nurseries to universities” - Investment 4.1 “Extension of the number of research doctorates and innovative doctorates for public administration and cultural heritage” - CUP E66E23000110001. Some of the results in this paper have been derived using the HEALPix~\cite{ag} package. Computational resources provided by INDACO Platform, which is a project of High Performance Computing at the Università degli Studi di Milano.\\

\centering
\includegraphics[width=15cm]{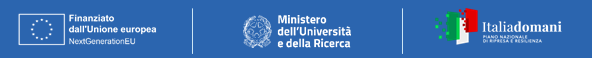} 

\vspace{10pt}
\includegraphics[width=5cm]{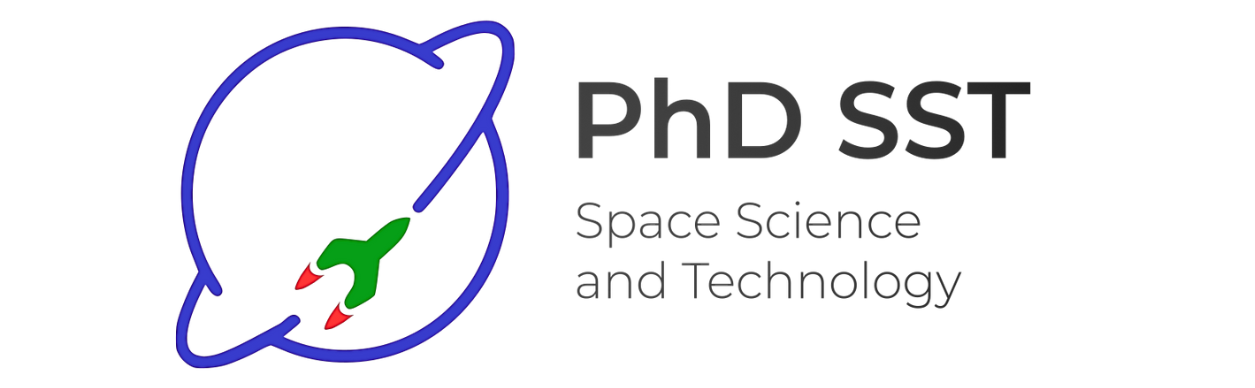}\hfill
\includegraphics[width=5cm]{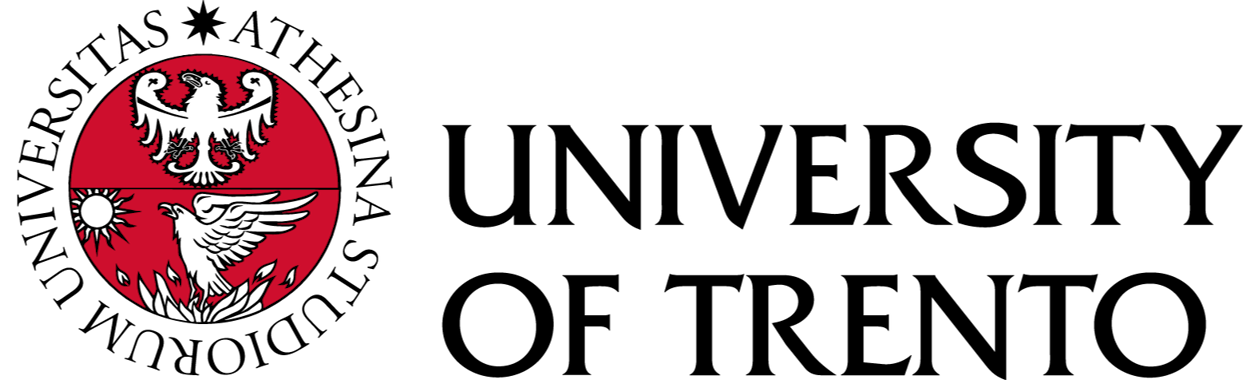}\hfill
\includegraphics[width=5cm]{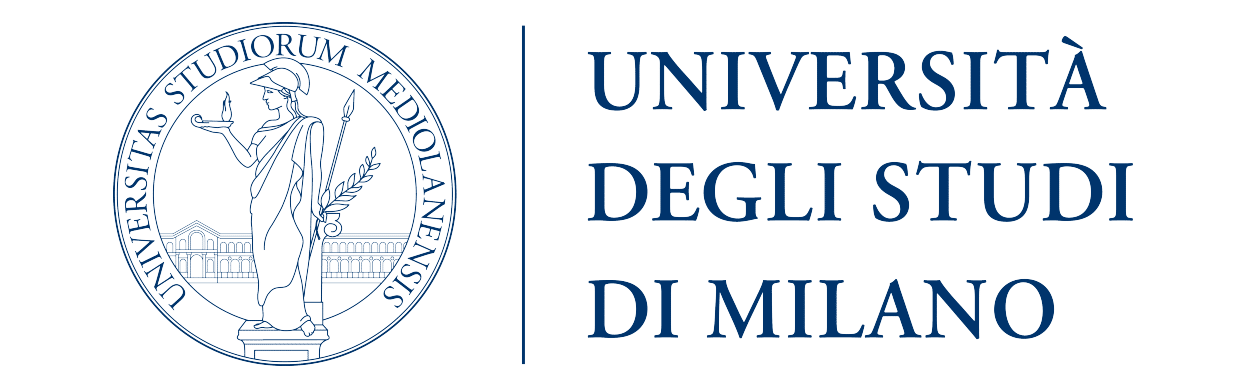}


\end{document}